\documentclass[conference]{ndss}
\usepackage{xspace}
\usepackage{color}
\usepackage{paralist}
\usepackage{wrapfig}
\usepackage{multirow}
\usepackage{listings}
\usepackage{makecell}
\usepackage{boxedminipage}
\usepackage{booktabs}
\usepackage{subcaption}
\usepackage{caption}
\usepackage{multirow}
\usepackage{multicol}
\usepackage{diagbox}
\usepackage{adjustbox}
\usepackage{xcolor}
\usepackage{url}
\usepackage{tabularx}

\usepackage{amssymb}
\usepackage{pifont}
\usepackage{cite}
\usepackage{textcomp}
\usepackage{balance} 
\usepackage{amsmath}
\usepackage{graphicx}
\usepackage{upquote}
\usepackage[breaklinks,colorlinks]{hyperref}
\usepackage{breakurl} 
\usepackage{amsthm} 
\usepackage[T1]{fontenc}

\usepackage{algorithm}
\usepackage{algpseudocode}

\usepackage{pifont}
\usepackage{balance}
\usepackage[htt]{hyphenat}

\newcommand{\bheading}[1]{{\vspace{2pt}\noindent{\textbf{#1}}}}
\newcommand{\iheading}[1]{{\vspace{2pt}\noindent{\textit{#1}}}} 

\newcounter{note}[section]

\newcommand{\etal}{\emph{et al.}\xspace}

\newcommand{\ie}{\emph{i.e.}\xspace}
\newcommand{\eg}{\emph{e.g.}\xspace}

\newcommand{\secref}[1]{\mbox{Sec.~\ref{#1}}\xspace}

\newcommand{\figref}[1]{\mbox{Fig.~\ref{#1}}}
\newcommand{\tabref}[1]{\mbox{Table~\ref{#1}}}

\newcommand{\ignore}[1]{}

\newcounter{packednmbr}

\newenvironment{packeditemize}{
\begin{list}{$\bullet$}{
\setlength{\labelwidth}{8pt}
\setlength{\itemsep}{0pt}
\setlength{\leftmargin}{\labelwidth}
\addtolength{\leftmargin}{\labelsep}
\setlength{\parindent}{0pt}
\setlength{\listparindent}{\parindent}
\setlength{\parsep}{0pt}
\setlength{\topsep}{3pt}}}{\end{list}}

\newcommand{\tx}{transaction\xspace}
\newcommand{\Tx}{Transaction\xspace}
\newcommand{\txs}{transactions\xspace}
\newcommand{\Txs}{Transactions\xspace}

\newcommand{\pri}{private\xspace}
\newcommand{\Pri}{Private\xspace}

\newcommand{\nalgref}[1]{\mbox{Alg.~\ref{#1}}}

\newcounter{lessoncount}

\newcommand{\lesson}[1]{
\refstepcounter{lessoncount}
\vspace{5pt}
\setlength\fboxrule{0.8pt}
\noindent\fbox{%
\parbox{0.96\linewidth}{%
   \textbf{Finding~\thelessoncount:} {#1}
}}}

\newcommand{\swap}{\textit{Swap}\xspace}

\hypersetup{
    colorlinks = true,
    citecolor = blue
}

\begin{document}



\title{An Empirical Study on Ethereum \\Private Transactions and the Security Implications}

\IEEEoverridecommandlockouts
\author{\IEEEauthorblockN{Xingyu Lyu$^1$\thanks{The first two authors contributed equally to this work.}~, 
Mengya Zhang$^2$$^*$,
Xiaokuan Zhang$^3$$^4$, 
Jianyu Niu$^1$,
Yinqian Zhang$^1$\thanks{Yinqian Zhang is affiliated with Research Institute of Trustworthy Autonomous Systems and Department of Computer Science and Engineering, Southern University of Science and Technology, Shenzhen, Guangdong, China.}~,
Zhiqiang Lin$^2$
}
\IEEEauthorblockA{$^1$Southern University of Science and Technology}
\IEEEauthorblockA{$^2$The Ohio State University}
\IEEEauthorblockA{$^3$George Mason University}
\IEEEauthorblockA{$^4$Georgia Institute of Technology}

}

\maketitle

\newcommand{\refappendix}[1]{\hyperref[#1]{Appendix~\ref*{#1}}}
\begin{abstract}
Recently, Decentralized Finance (DeFi) platforms on Ethereum are booming, and numerous traders are trying to capitalize on the opportunity for maximizing their benefits by launching front-running attacks and extracting Miner Extractable Values (MEVs) based on information in the public mempool. 
To protect end users from being harmed and hide \txs from the mempool,  {\it private transactions}, a special type of transactions that are sent directly to miners, were invented.
\Pri \txs have a high probability of being packed to the front positions of a block and  being added to the blockchain by the target miner, without going through the public mempool,
thus reducing the risk of being attacked by malicious entities.

Despite the good intention of inventing private \txs,
due to their stealthy nature, private \txs have also been used by attackers to launch attacks,
which has a negative impact on the Ethereum ecosystem.
However, existing works only touch upon private \txs as by-products when studying MEV, while a systematic study on \pri \txs is still missing.
To fill this gap and paint a complete picture of private \txs,
we take the first step towards investigating the private \txs on Ethereum.
In particular, we collect large-scale private \tx datasets and perform analysis on their characteristics, transaction costs and miner profits, as well as  security impacts.
This work provides deep insights on different aspects of \pri \txs.
\end{abstract}

\section{Introduction}
\label{sec:intro}

Recent years have witnessed an explosive growth of DeFi~\cite{defi}, which provides end users with financial products and services. By June 2022, the Total Value Locked (TVL) of DeFi has reached about \$114 billion~\cite{dappradar}. Moreover, the number of DeFi wallets has increased to around \$4.8 million as of May 2022~\cite{consensys}.
Among all blockchains, Ethereum is the second-largest blockchain by market capitalization and the first to support smart contracts, which are the foundation of DeFi. Nowadays, the majority of DeFi TVL (\$91 billion) lies in Ethereum. 

Due to the popularity of DeFi platforms and the large amount of money involved, the number of attacks to steal money from them is also arising.
Since all \txs in Ethereum need to be broadcasted before mining,
every \tx will need to stay in the public mempool for some time.
Some attackers have exploited this fact and launched attacks (\eg, frontrunning~\cite{frontrunning}) targeting pending \txs in mempool.
In particular, in frontrunning attacks,
an attacker observes a victim \tx in the mempool and launches an attack \tx with certain features (\eg, higher gasprice)
so that the attack \tx will be mined before the victim \tx, thus making profits. 
Another example is the Miner Extractable Value (MEV)~\cite{flashboy},
which represents the profit that miners can extract from the manipulation of transactions. For example, two MEV Bots have gained \$476,000 recently by targeting the stablecoin swaps~\cite{mevbot}. 
\looseness=-1

\begin{figure*}[t]
\centering
\includegraphics[width=.9\linewidth]{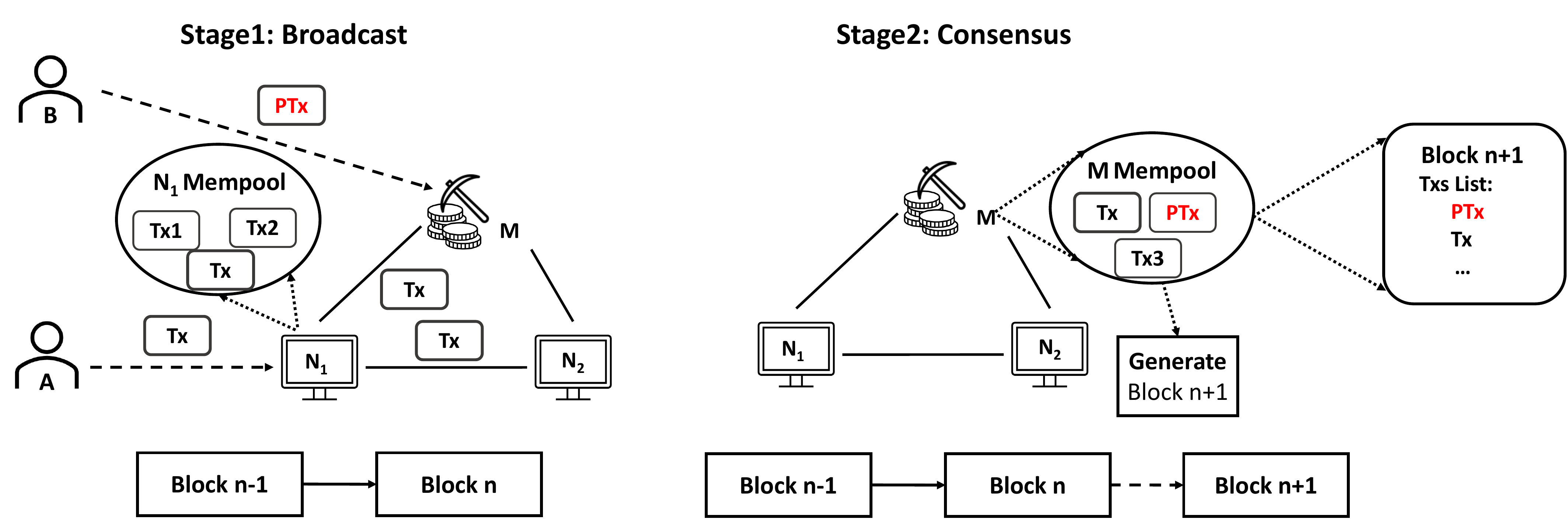}
\caption{The two stages of mining a normal transaction (\texttt{Tx}) vs. a \pri \tx (\texttt{PTx}) in Ethereum. 
User \texttt{A} sends a normal transaction (\texttt{Tx}) to a node ($N_1$), which is then put into the mempool of this node. Generally, in Stage 1 $N_1$ broadcasts \texttt{Tx} to its peers, $N_2$ and $M$ (M is the miner). 
However, the Stage 1 is different for a \pri \tx (\texttt{PTx}) sent from User \texttt{B}, in which \texttt{PTx} is sent directly to $M$.
In Stage 2, $M$ solves the Proof-of-Work puzzle, and includes \texttt{PTx} and \texttt{Tx} in the newly mined block. 
In particular, \texttt{PTx} is put in front of \texttt{Tx}.}
\label{figs:bg}
\end{figure*}

To help protect \txs from being attacked, \textbf{\pri \txs} were proposed. 
A private transaction is a special transaction that 
can be sent directly to miners, 
bypassing the public mempool (see~\figref{figs:bg}). 
By doing so, such \txs remain private (\ie, only present in the target miner's mempool)
until they are posted by the target miner, 
and cannot be monitored by others. 
As a result, attackers cannot see these \txs in their mempool, 
thus thwarting the attacks. 
Besides being hidden from others,
\pri \txs have a high probability 
of being packaged to the front position of a new block 
by the miner and then added to the blockchain. 
This property also motivates MEV searchers, 
frontrunning attackers, and arbitragers
to seek profit-making opportunities.


It has been two years since \pri \txs were first introduced by \textit{SparkPool} in August 2020~\cite{first-ptx}.
However, private \txs have not attracted much attention from the research community.
There are several works that have analyzed the Ethereum blockchain and DeFi platforms,
but they mainly focus on 
1) detecting bugs from smart contracts~\cite{ luu2016making, nikolic2018finding, tsankov2018securify, krupp2018teether, kalra2018zeus,torres2018osiris,frank2020ethbmc,torres2021confuzzius},
2) measuring Ethereum networks and \txs~\cite{lee-measure, chen2020understanding, Zhao2021TemporalAO, bai-temporal, said-tx, lin-tx, zanelatto2020transaction}, and
3) analyzing \txs to uncover attacks~\cite{Zhang2020TXSPECTORUA, wu2021defiranger, sereum-ndss19, qin2021attacking}.

To date, however, there are only a few papers~\cite{piet2022extracting,weintraub2022flash,qin2021quantifying,capponi2022evolution} that touched upon \pri~\txs as by-products when they study MEVs. However, the scope of \pri~\txs goes beyond MEVs. A complete view of private \txs such as their characteristics and impacts on the Ethereum ecosystem remains unclear.
Although the original intention of inventing \pri \txs is to protect users from attacks, how they are actually used in reality is still an open question.


In this paper, we make the first step towards understanding \pri~\txs and their impacts on the Ethereum ecosystem.
Specifically, we conduct a large-scale empirical study on \pri~\txs from three different dimensions: 1) the basic characteristics of \pri~\txs, 2) their economic impacts to Ethereum such as  \tx cost and miner profits of \pri\txs, and 3) their security implications such as the real-world attacks hidden in \pri\txs. 
%
%
We collect \tx related information from customized Ethereum nodes, 
and we retrieve public data from Etherscan~\cite{etherscan}, TradingView~\cite{tradingview}.
%
In total, we collect four datasets for our analysis, 
containing transaction data from May 1, 2021 to April 30, 2022 (one year)
and mempool observation data from May 22, 2022 to May 30, 2022 (nine days).
Based on these large-scale datasets, we perform a detailed analysis of the basic characteristics of \pri \txs, their impacts on economics as well as on security.   To the best of our knowledge, we are the first to systematically study the \pri \txs and their impacts on the blockchain ecosystem. 

\bheading{Contributions.} 
This paper makes the following contributions:
\begin{packeditemize}
\item \bheading{Large-scale \pri \tx datasets (\secref{sec:dataset}).}
We construct four datasets
from public data sources and our modified Ethereum nodes. 
The datasets include transaction information in a one-year timespan,
as well as the DeFi platforms, miners, ERC20 tokens, and MEV Bots involved in each of the \txs.
We also build datasets from the local mempools
of two Ethereum nodes deployed on two continents,
which is used to measure private \tx leakage.
To facilitate future research, 
all the datasets used in our paper will be open-sourced.

\item \bheading{Characteristics of \pri \txs (\secref{sec:basic}).}
We study the characteristics of \pri \txs. 
We find that during the last year, 
the percentage of \pri\txs per month is increasing 
and rises to about 2\% of the total \tx volume. 
We categorize the purpose of \pri \txs. 
Although \pri \txs were proposed to protect end users from attacks, 
we find that only 18.1\%  of them were used for that purpose,
whereas 28.6\% of them are actually related to MEV Bots.
Besides, five of the top ten receivers of \pri~\txs are MEV Bots.

\item \bheading{Impacts of \pri \txs on blockchain economics (\secref{sec:money}).}
We study how \pri \txs affect the transaction fees and miner profits.
We find that the gas used for private transactions is about
737,829 on average, which is much smaller than the
average gas used (16,673,757) of normal transactions.
We also investigate the impact of EIP-1559 on \pri \txs. 
We find that around 50\% of \pri \txs set the gas price at zero before EIP-1559; 
after EIP-1559, since basefees are mandatory, 
there are around 22\% private transactions setting the priority fee to zero. 
The revenue of \pri \txs is an integral part of miner profits, 
accounting for around 5\% of the total revenue. \looseness=-1

\item \bheading{Impacts of \pri \txs on blockchain security (\secref{sec:security}).}
We study security issues related to \pri \txs, 
including MEV, attack case studies, 
consensus security and leakage of \pri \txs.
We find that 2.6\% \pri \txs senders 
earned more than ten ETH as profits via MEV Bots.
Attackers have already utilized \pri \txs to launch attacks~\cite{lifi-attack, mutlichain-attack, multichain-return};
in these attacks,
the attackers paid a large amount to the miner as a bribe 
to get their \txs executed.
According to our evaluation,
the miner earned as high as 700 ETH for mining a single \pri \tx. 
This can lead to serious consensus security issues, 
such as the undercutting attacks~\cite{carlsten2016instability, flashboy}.
We also find that \pri \txs are not always private. 
By running two Ethereum nodes in two continents for nine days, 
we have observed 4.3\% \pri \txs in our mempool,
which means that they are actually not private. 
Therefore, users should proceed with caution when sending \pri \txs.
\end{packeditemize}

\ignore{
\bheading{Roadmap.} 
The remainder of the paper is outlined as follows. 
\secref{sec:bg} introduces the background.
\secref{sec:related} discusses the related works. 
\secref{sec:sys} explains the datasets used in the paper.
\secref{sec:eval} analyzes the characteristics of \pri \txs.
\secref{sec:money} measures the transaction fee and miner profits. 
\secref{sec:security} presents the security related issues, including MEV and real-world attacks.
In the meanwhile, we discuss implications and limitations of our study in \secref{sec:discuss} and conclude the paper in \secref{sec:conclude}. 
}


\section{Background}
\label{sec:bg}

\subsection{Ethereum Basics}
Ethereum~\cite{Buterin2013} is a public, decentralized, and permissionless blockchain platform, which supports Turing-Complete smart contracts. In Ethereum, nodes, also called Ethereum clients, are connected through a P2P network. Moreover, nodes run a discovery protocol to find other nodes~\cite{discovery} and a TCP-based transport protocol for data communication~\cite{tcp}. 
In particular, the nodes who have the capabilities to create new blocks are called miners. To produce a new block, a miner has to pack \txs into the block and then solve the difficult puzzle according to the Proof-of-Work (PoW)~\cite{pow-web} consensus mechanism. 
Once mined, the miner broadcasts the block to other nodes for validation and execution. 
For each new block, miners can receive block rewards and gas fees (introduced shortly) as earned profits. 
There is a gas limit of the total gas used by the \txs within one block. 
Note that Ethereum 2.0 will use Proof-of-state (PoS) to replace PoW in the near future~\cite{pos}.

\bheading{ERC20 tokens.}
Ethereum Request for Comments 20 (ERC20)~\cite{erc20} is a token standard for fungible tokens, which is the second most popular token type in Ethereum, in addition to \textit{ETH}.
Any user can own and even create such tokens by implementing the required smart contract features,
such as \textit{transfer()}, \textit{transferFrom()} and \textit{balanceOf()} functions, as well as related Events.

\subsection{DeFi}
ERC20 tokens are the most popular tokens created and used by DeFis.
DeFis are decentralized financial applications consisting of multiple smart contracts as backends.  
DeFis mainly serve for borrowing and lending, token exchanging, asset management, and other similar financial functionalities.
By February 2022, there are around 4.4 million unique Ethereum addresses that have interacted with DeFi contracts~\cite{report}.

\bheading{Stablecoin.}
Stablecoins are implemented based on the ERC20 token standards to ensure the price stability. 
For example, Tether (USDT)~\cite{tether} is a stablecoin with a price pegged to 1 USD.
Similarly to other types of ERC20 tokens, they are global to reach in the Ethereum network and exchangeable with any other tokens.
As the time of writing, the top three stablecoins by market capitalisation~\cite{scoin} are Tether, USD Coin, and TerraUSD. 

\subsection{Gas and Fees}
To pay for the computational resources to execute a \tx, every \tx is required to pay a fee that is decided by both the gas and the gas price. Specifically, the \tx fee is calculated as: \textit{TxFee = UsedGas × GasPrice},
where \textit{UsedGas} refers to the gas amount used for executing a \tx 
and \textit{GasPrice} is the amount that the user would like to pay per unit of gas. 
As previously mentioned, a gas limit is enforced to restrict the total gas used of \txs in each block. 

\bheading{EIP-1559.}
EIP-1559~\cite{eip1559} is a proposal requiring \txs to pay both the base fee and the priority fee as the total gasprice.
Specifically, basefee is algorithmically determined for each block in Ethereum, and the priority fee is the optional fee to motivate miners to include the \txs in new mined blocks.
Before EIP-1559, there is no limitation on the gasprice. Users can set zero to the gasprice of their \txs. However, after the EIP-1559 taking effects, the gasprice is required to be equal or higher than the basefee of the mined blocks. On the other hand, the basefee is the minimum gasprice for mined \txs. 
Generally, miners usually choose the \txs with the higher gas price, since the higher the gasprice, the more profits miners will earn. 
Besides, to avoid network congestion, the gas limit of each block is doubled from 15,000,000 \textit{GWei} to 30,000,000 \textit{GWei}.


\subsection{Private Transactions}
\bheading{\Pri~\txs.}
Every normal \tx has to go through two stages to be mined in blockchain, as shown in~\figref{figs:bg}. 
In particular, a user sends a \tx to Ethereum nodes in the p2p network, and all nodes will add this \tx to their local mempool. 
In contrast, a \pri \tx is sent directly to the miners and will not appear in the mempool of normal nodes. 
Then, miners pack these \txs into blocks and put the \pri \tx in front of the normal \tx. 
However, there is a price of being a \pri \tx, which is usually to pay the additional fee by directly sending money during the execution of \tx. 
Moreover, the \pri \tx can be leaked as discussed in~\cite{ptx-leakage}. 
We perform experiments to confirm the leakage and discuss the related security issues in~\secref{sec:security:leak}.

\bheading{Private \tx relay systems.} A relay system is built to provide a private channel for \txs directly being submitted from users to miners. Flashbots~\cite{flashbot} are one of such relay systems, and around 90\% of miners use Flashbots to earn such extra profits~\cite{flashbot2}. Firstly, an Ethereum user that is usually a searcher, searches for the MEV opportunities to submit \pri \txs and such \txs are then packed into a bundle. Secondly, relayers that are bundle propagation services forward the \txs to miners. Finally, miners mine the \txs into the front places of the new mined block as long as they are satisfied with the rewards. The workflow is similar in other relay systems, such as Eden Network~\cite{edennetwork}.

\subsection{MEV}
\textit{Daian et al.}~\cite{flashboy} first propose the concept of MEV and introduce the potential risks brought by MEV. 
Although MEV is called miner extractable value, it is usually users, instead of miners, that search for MEV opportunities and share the earned profits with miners. Miners only need to put MEV related \txs to the top places of blocks and take shared profits from users.
To compete for extracting MEV, MEV Bots are created to monitor the blockchain network automatically for transactions with potential MEV opportunities and immediately launch related \txs via channels (e.g., Flashbots~\cite{flashbot}). 
Since being coined in 2019, MEV has been intensively studied by researchers~\cite{piet2022extracting,weintraub2022flash}.
The extensive exploration on MEV might cause instabilities and re-organization on the blockchain network, since miners are incentivized to re-mine blocks. 
Moreover, users might suffer financial losses if they are attacked or front-run by MEV \txs. 

\section{Dataset}
\label{sec:sys}
\label{sec:dataset}

\begin{table*}[]
\centering
\footnotesize
\begin{tabular}{l|llll}
\toprule
Dataset & Timespan & Data Source & Information & Used In \\\hline
One-year Replayed Transaction & One year & Geth Replay & Block info, Transaction info &\secref{sec:basic};~\secref{sec:money};~\secref{sec:security} \\
Nine-day Mempool Transaction & Nine days & Geth Mempool & Tx info received in mempool &\secref{sec:security} \\
Smart Contract Label & One year & Etherscan, TradeView & Miner, MEVBot, DeFi, Token &\secref{sec:basic};~\secref{sec:money};~\secref{sec:security} \\
Private Transaction Label & One year + Nine days & Etherscan & Private transactions &\secref{sec:basic};~\secref{sec:money};~\secref{sec:security} \\ \bottomrule
\end{tabular}
\caption{Datasets used in the paper. One-year timespan is from May 1, 2021 to April 30, 2022; Nice-day timespan is from May 22, 2022 to May 30, 2022.}
\label{tab:datasets}
\vspace{-10pt}
\end{table*}

To empirically study \pri \txs, we extract the necessary data from some reliable data sources including Etherscan Label Word Cloud~\cite{etherscan-label} and TradingView~\cite{tradingview}, as well as our modified Ethereum nodes. 
In this section, we explain how we collect data in detail.
The datasets and how they are used in later sections are presented in \tabref{tab:datasets}.

\subsection{One-year Replayed Transaction Dataset}
\label{sec:dataset:tx}
We collect the necessary \tx and block information from our customized \textit{Geth}\cite{geth} node in full mode, which is an official Ethereum client implemented in Go language. 
Specifically,
we replay every \tx from May 1, 2021 (block 12,344,945) to April 30, 2022 (block 14,688,626) to extract information and construct our one-year dataset, which contains 446,925,956 \txs in total. We modify the \textit{Geth} node to collect the following information:
    
\bheading{Block information.} 
For every block, the obtained information includes block number, timestamp, miner address, block reward, gas limit, gas used, basefee, and burnt fee. 
    
\bheading{Transaction information.} 
For each \tx, we first collect the basic information, including \textit{hash}, block number, success status (0 means failure and 1 means success), sender address, receiver address, ETH value, input, gasused, gasprice, and gasfee.
We also collect the following information:
 
 \iheading{1) Traces}: the information of \textit{internal transactions} that are the smart contract calls within the \tx. Specifically, every \textit{internal transaction} should record the sender address, receiver address, ETH value, input, and  the call graph information. 
 
  \iheading{2) Money flows}: token amount, token address, token sender, token receiver, and \textit{MFGIndex}. In particular, \textit{MFGIndex} records the index of the position of current money flow among all the money flows that include both \textit{ETH} and ERC20 token transfers.
 
 \iheading{3) Money flow graph (MFG)}: 
 To analyze the used tokens and DeFi platforms, for every DeFi-related \tx, we build a money flow graph (MFG). In MFG, nodes represent the token sender and receiver, and edges indicate the token related information, including \textit{MFGIndex}, token name, and token amount.

\subsection{Nine-day Mempool Transaction Dataset}
\label{sec:dataset:mempool}
To measure the leakage of \pri \txs and see if we can observe any of them before they are mined on blockchains, we deploy two modified Ethereum nodes in two continents from May 22, 2022, to May 30, 2022 (nine days) and collect the received \txs from the local mempool. Note that \txs in the mempool are not yet mined.
Specifically, we customized the \textit{Geth} node to log the hash, block number, timestamp of \txs observed from the mempool of the two nodes.  We obtained 6,720,710 \txs from Node 1 and 7,854,054 \txs from Node 2 during the nine days.


\subsection{\Pri \Tx Label Dataset}
\label{sec:dataset:private-tx}
We obtain the \pri \txs within both the one-year dataset and the nine-day dataset by crawling Etherscan Label Cloud~\cite{etherscan-label}. 
Specifically, we observe 7,405,835 \pri \txs obtained from our one-year \tx dataset and the remaining 439,520,121 are normal (non-private) \txs.
We also obtain 232,729
 \pri \txs from the nine-day dataset. 

\subsection{Smart Contract Label Dataset}
\label{sec:dataset:sc}
To measure the behaviors of different entities involved in the \pri \txs, for each address in our one-year dataset, we check whether they have the following labels; if so, we collect the corresponding information:

 \iheading{1) Miner}: miner address and the related label. We collect the miner address of every block in our dataset and check the address related label\footnote{e.g., {\scriptsize 0xEA674fdDe714fd979de3EdF0F56AA9716B898ec8 is labeled as \textit{Ethermine}}.} from the Etherscan Label Cloud.
 
 \iheading{2) MEVBot}: addresses whose label are MEVBots. For every unique sender and receiver of \txs in our dataset, we check its label from Etherscan. If the label is MEVBot, we collect it.
 
 \iheading{3) DeFi}: DeFi platform address and the related label. Similarly, we collect the addresses which are marked as DeFi by Etherscan, and their labels. 
 
 \iheading{4) Token}: For each smart contract labeled as ERC20 token, we obtain the token address and name from Etherscan, and historical price data from TradingView.

\section{Characteristics}
\label{sec:basic}
In this section, we describe the general statistics of \pri \txs by measuring their categories classified by their purposes, the involved DeFi tokens and platforms, and the involved entities.

\begin{figure*}
\centering
\begin{minipage}[b]{.32\linewidth}
  \centering
  \includegraphics[width=\linewidth]{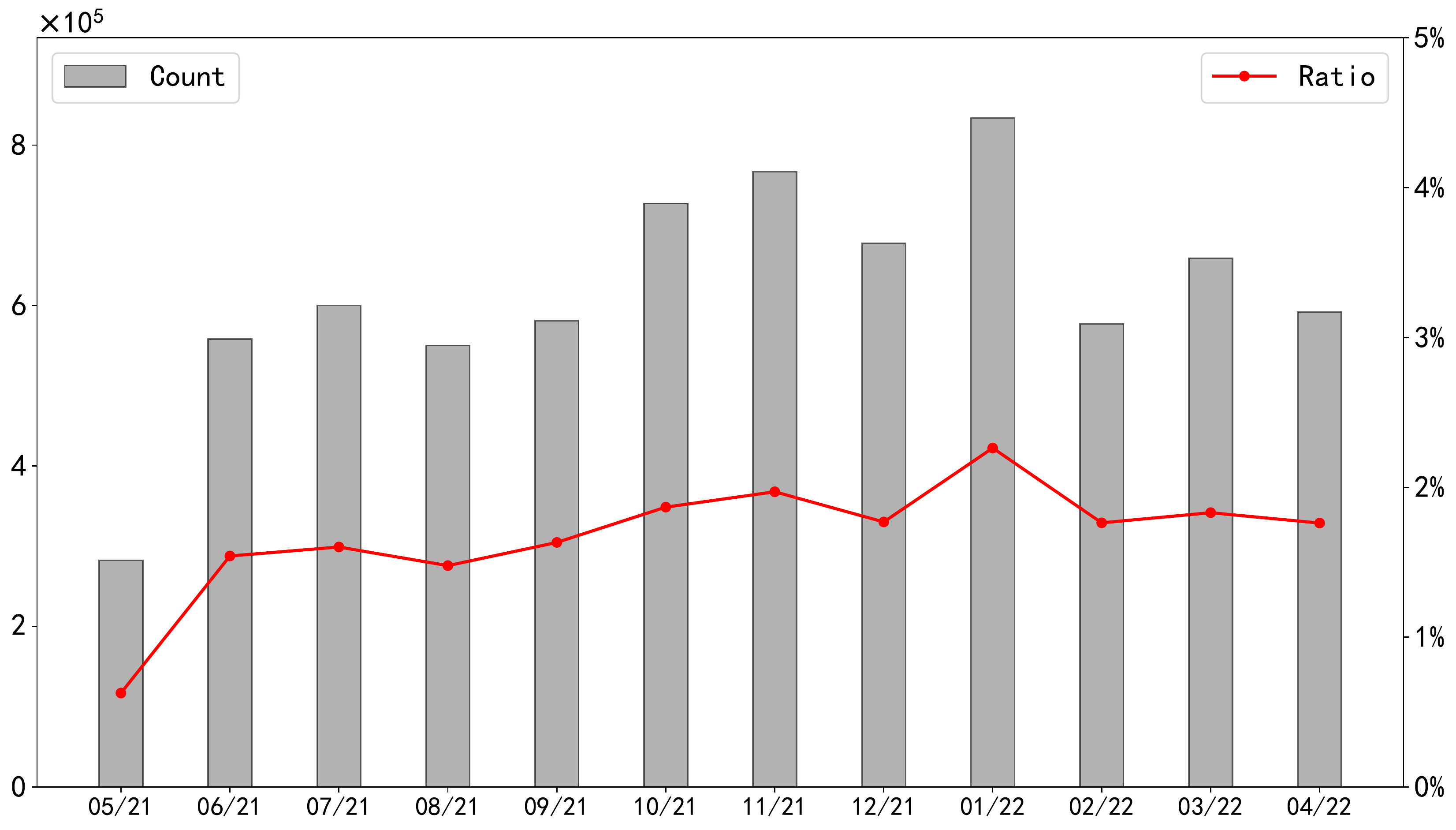}
  \captionof{figure}{The count and percentage of private transactions per month.} 
    \label{figs:txnum}
\end{minipage}%
\hfill
\begin{minipage}[b]{.32\linewidth}
  \centering
  \includegraphics[width=\linewidth]{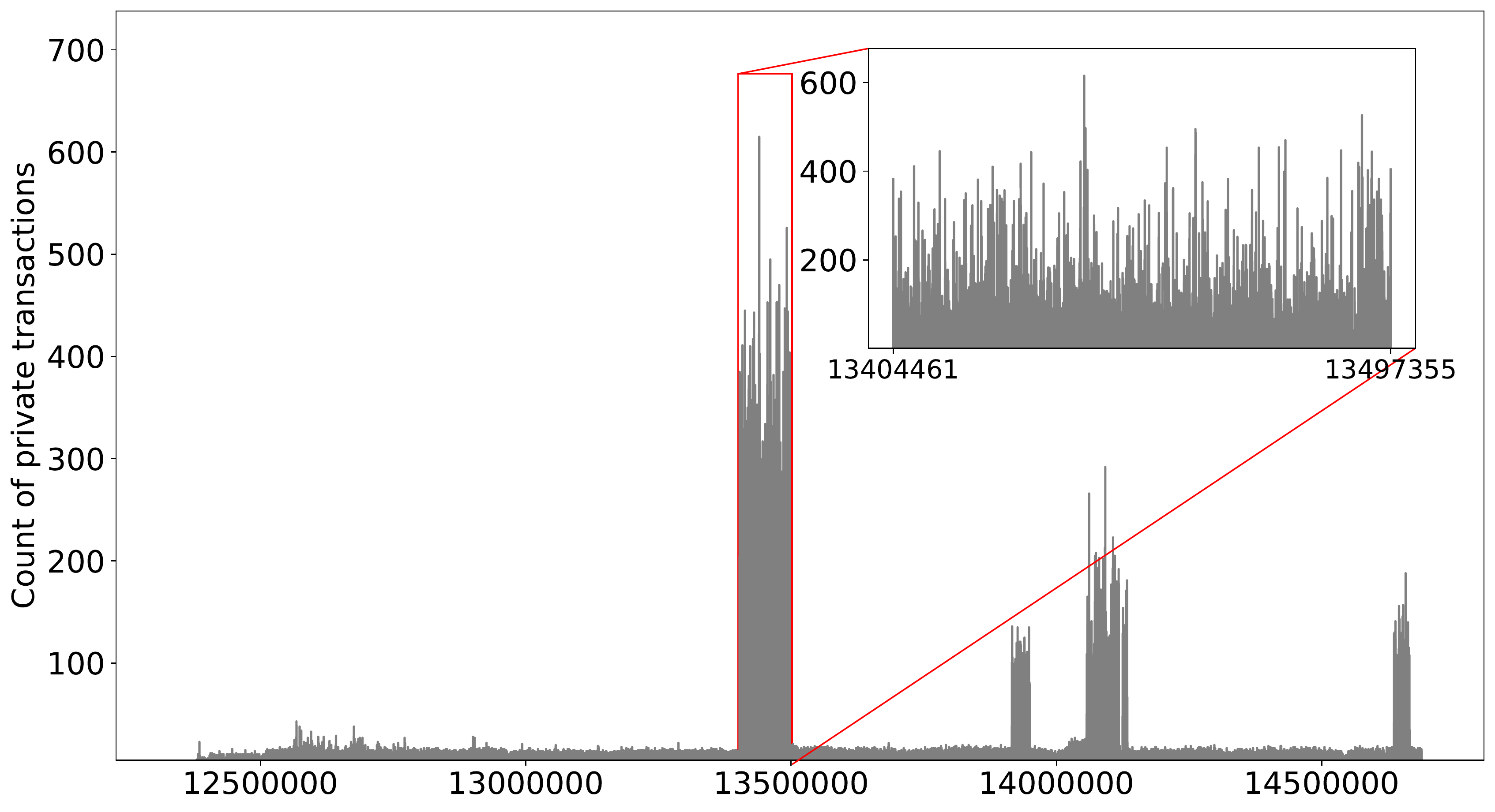}
  \captionof{figure}{The count of \pri \txs per block.} 
    \label{figs:pritxperblock}
\end{minipage}
\hfill
\begin{minipage}[b]{.32\linewidth}
  \centering
  \includegraphics[width=\linewidth]{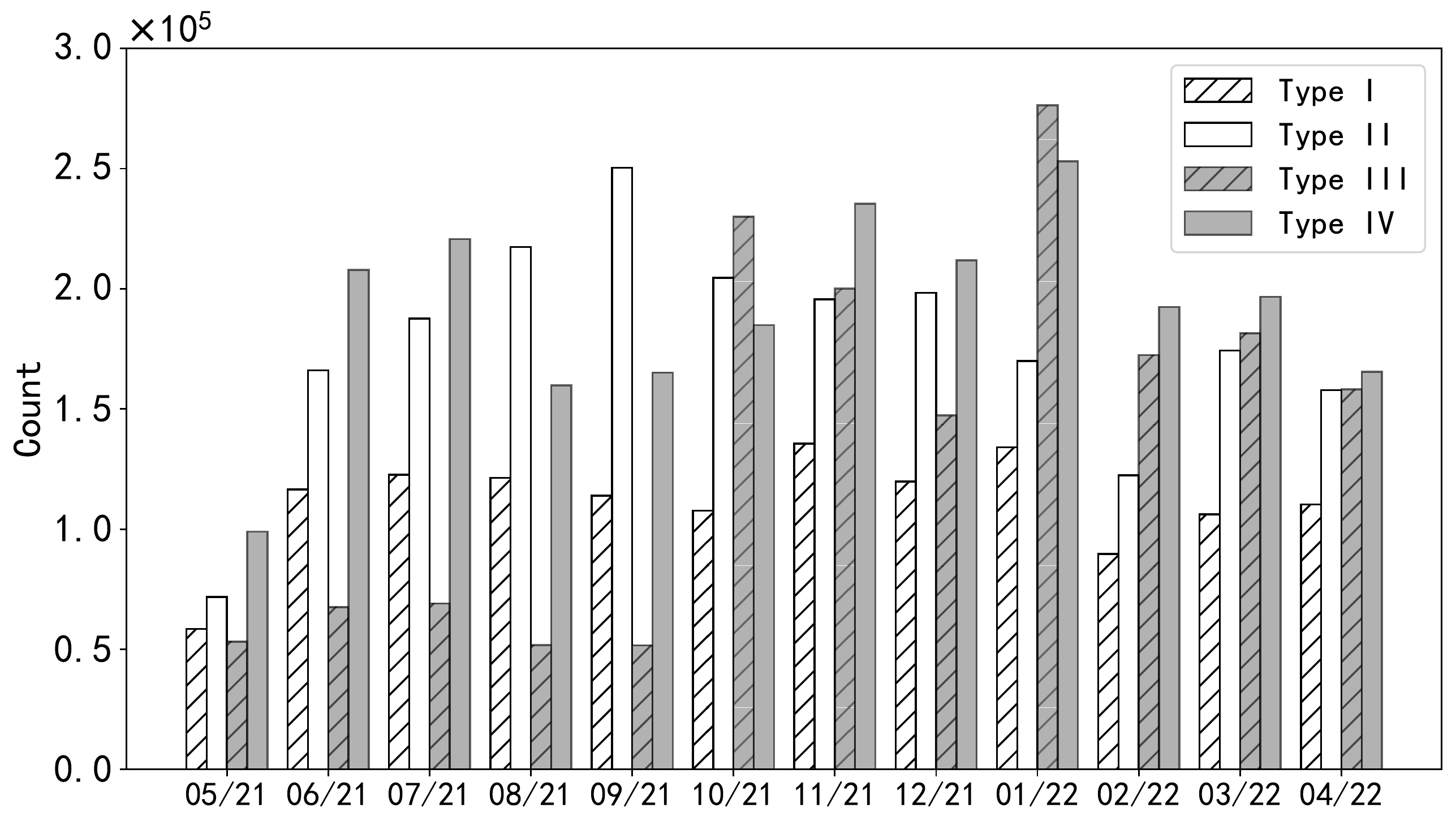}
  \captionof{figure}{The count of each type of \pri \txs per month.}
    \label{figs:typepermonth}
\end{minipage}
\end{figure*}


\subsection{Distribution}
\bheading{The count and percentage of \pri \txs.} 
    \figref{figs:txnum} shows the count of \pri \txs against per month.
    There are at least 0.28 million \pri \txs every month and around 0.60 million \pri \txs on average per month. 
    Besides, the percentage of \pri \txs among all the \txs account for 2\% on average per month. 
    Overall, there are not many \pri \txs in the total mined \txs, while the percentage of \pri \txs keeps increasing slightly every month. 

    
\bheading{The distribution of \pri \txs.}
\label{sec:char:dis}
    \begin{figure}[t]
    \centering
    \includegraphics[width=1\columnwidth]{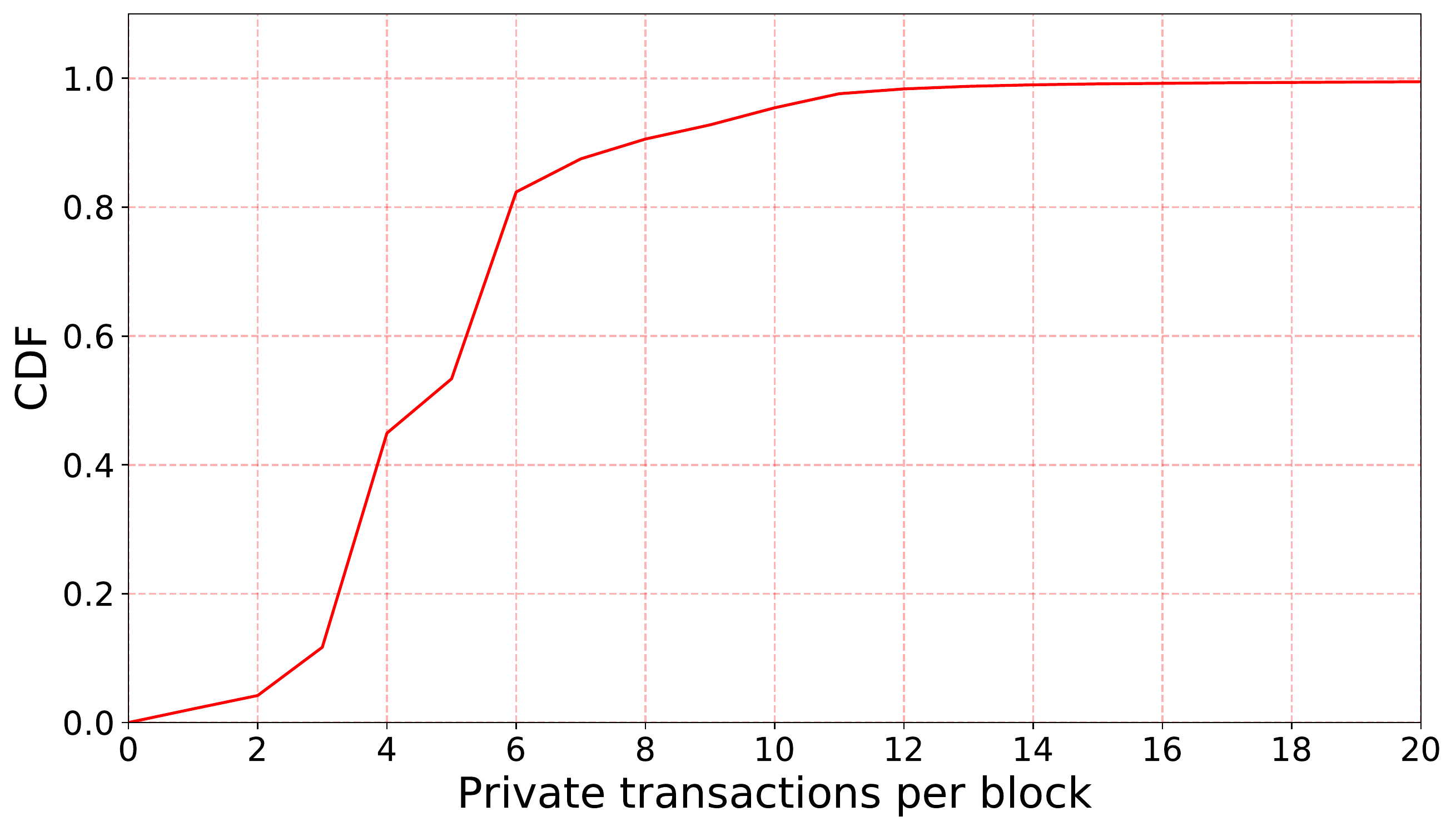}
    \caption{Number of \pri \txs per block. Only 0.5\% of them have more than 20 \pri \txs.} 
    \label{figs:rate_mix_perblock}
    \vspace{-10pt}
    \end{figure}
    %
    %
     \figref{figs:rate_mix_perblock} plots the CDF of \pri \txs within a single block for the blocks that have at least one \pri \tx (1,259,251 blocks, 54\%).
    93.2\% of these blocks have less than 10 \pri \txs, and 
     99.5\% of these blocks have no more than 20 \pri \txs. Only 0.5\% have the count of \pri \txs greater than 20.
    The largest count of private \txs in one block is 615.

 
    \bheading{Abnormal count of \pri \txs in some blocks.}
    \figref{figs:pritxperblock} presents the count of \pri \txs against per block and also the blocks in the peek periods. There are four ranges of blocks that seem to have far more \pri \txs than other blocks. The block ranges are [13,404,461 - 13,497,355], [13,926,965 - 13,949,430], [14,057,403 - 14,133,216], and [14,636,264 - 14,664,418], respectively. 
    We show the details of the biggest peek in the figure.
    In particular, not all blocks have far more \pri \txs than others, and some blocks have the count of \pri \txs below the average. 
    In particular, we study the blocks with more \pri \txs than others and find that many \pri \txs are used for miners transferring \textit{ETH} to EOAs. Moreover, the other three peeks have similar patterns.
    
    

    \bheading{The positions of \pri \txs inside blocks.}
    \Pri \txs are usually bundled together to miners and placed at the top of the block. With the collected 1,259,251 blocks who have at least one \pri \tx, we observe that all \pri \txs are located in front of normal \txs in blocks. The results are expected since \pri \txs are created to race for the opportunities before other \txs. Thus, they should be put in the top positions of the blocks.  
    

\subsection{Categories}
\label{sec:char:cate}
    To better measure the usages and draw insights on the purposes of \pri \txs, we categorize them into the following four types.
    \begin{packeditemize}
        \item Type I (Self protection): the \pri \txs that are only used for DeFi services (e.g., Uniswap trader). Specifically, such \pri \txs protect themselves from risks in the public mempool, such as frontrunning attacks. 
        \item Type II (MEV related): the \pri \txs that MEV Bots participate in. \Pri \txs of this type should have MEV Bots as their recipients and are usually used for MEV extraction.
        \item Type III (Miner payout): the \pri \txs used to transfer money to EOAs. \Pri \txs of this type are normally sent from miners or other well-known services and used to transfer \textit{ETH}.
        \item Type IV (Normal services): the remaining \pri \txs whose receivers are normal smart contracts that cannot be labeled.
    \end{packeditemize}

\bheading{The categories of \pri \txs.}
    We present the count of \pri \txs of each type against per month in~\figref{figs:typepermonth}. 
    For type I, about 18.1\% of \pri \txs are sent to call DeFi related services. 
    For type II, there are 28.6\% of \pri \txs that are sent to MEV Bots, which shows the increasing need for \pri \txs for MEV searchers since the competition of MEV becomes more and more fierce.
    For type III, 22.4\% of \pri \txs are used for transfers. In particular, the senders of these \pri \txs are miners and well-known services(e.g., Crypto.com~\cite{crypto}).  
    In addition, the left 30.9\% of \pri \txs belong to type IV. Among them, about 650,000 transactions are used for transferring tokens through the smart contracts. 

   
    \lesson{Only 18.1\% of the total \pri \txs are used for the original purpose---protecting DeFi services from attacks. More \pri \txs (28.6\%) are used by MEV Bots in search of MEV opportunities. The actual usage of \pri \txs has departed from its design purpose.}
    

\subsection{DeFi Tokens and Platforms}
    Most \Pri \txs are used to interact with the tokens for different purposes mentioned in~\secref{sec:char:cate}.
    We first briefly describe some features of the tokens in \pri \txs, to give an overview of the usage of tokens. 
    Moreover, we classify the token exchange pattern into different types according to the token exchange count and token types.
    Then we demonstrate the usages of DeFis among the \pri \txs and the related tokens, for further deep analysis.

     \begin{figure}[t]
    \centering
    \includegraphics[width=1\columnwidth]{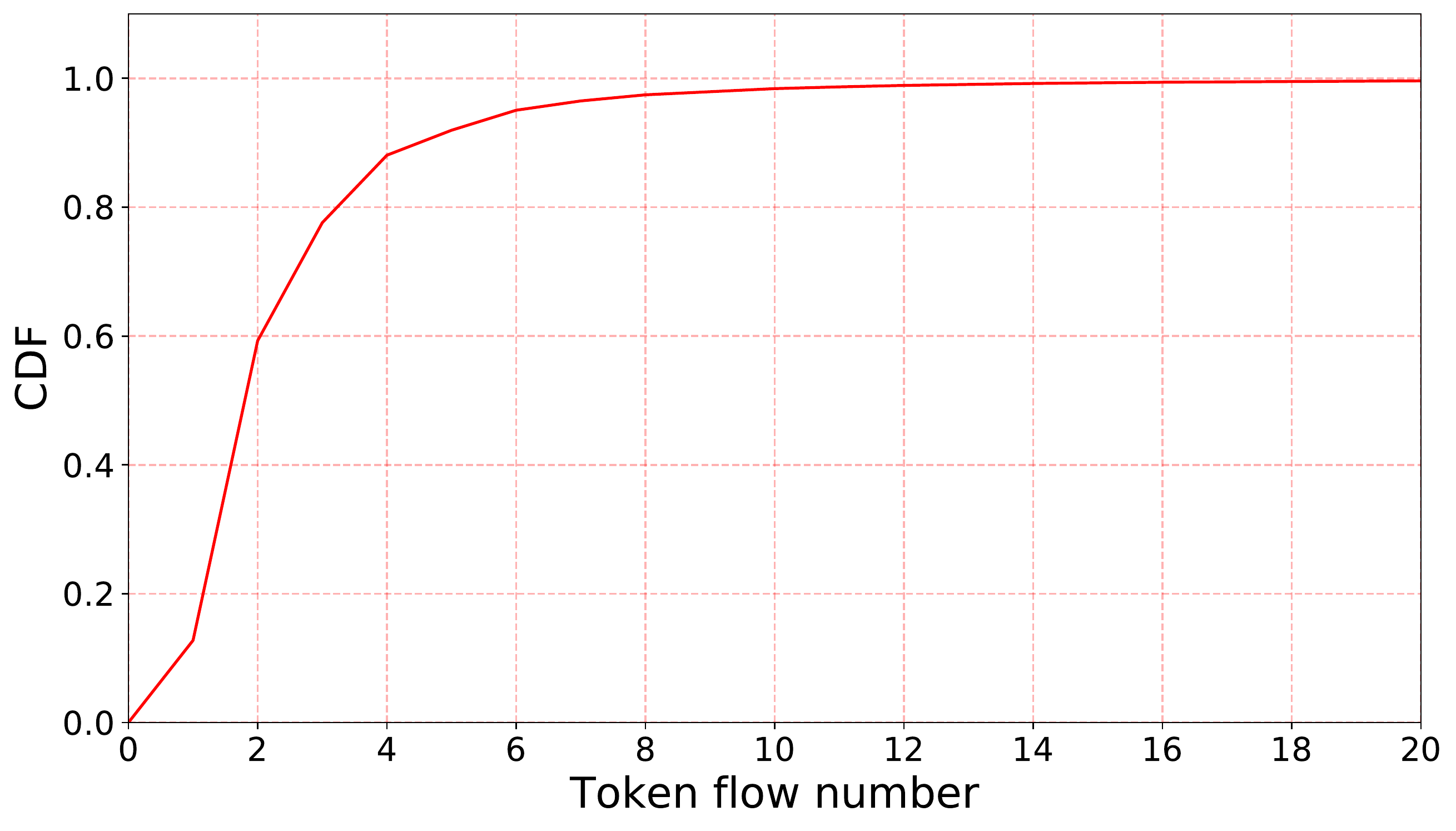}
    \caption{Number of token flows per \pri \tx. Only 0.4\% of them have more than 20 token flows.}
    \label{figs:token_flow}
    \end{figure}
    
   \bheading{The distribution of token flow number in \pri \txs.}
    We present the distribution of token flow numbers of \txs that have at least one and no more than 20 token flows (99.6\% of 5,746,633 \pri \txs) in~\figref{figs:token_flow}. We exclude 1,659,202 \pri \txs without token flow. From the figure, around 88\% \txs have no more than 4 token flows. In particular, the \pri \tx~\footnote{\scriptsize 0x470ceedf30f8c1c533b0911c2b37ab1c55b0cbd83d1e40c0d44cd7e9f5db1569} has the most token flows (600). 
    
\begin{figure}[t]
    \centering
    \includegraphics[width=.9\columnwidth]{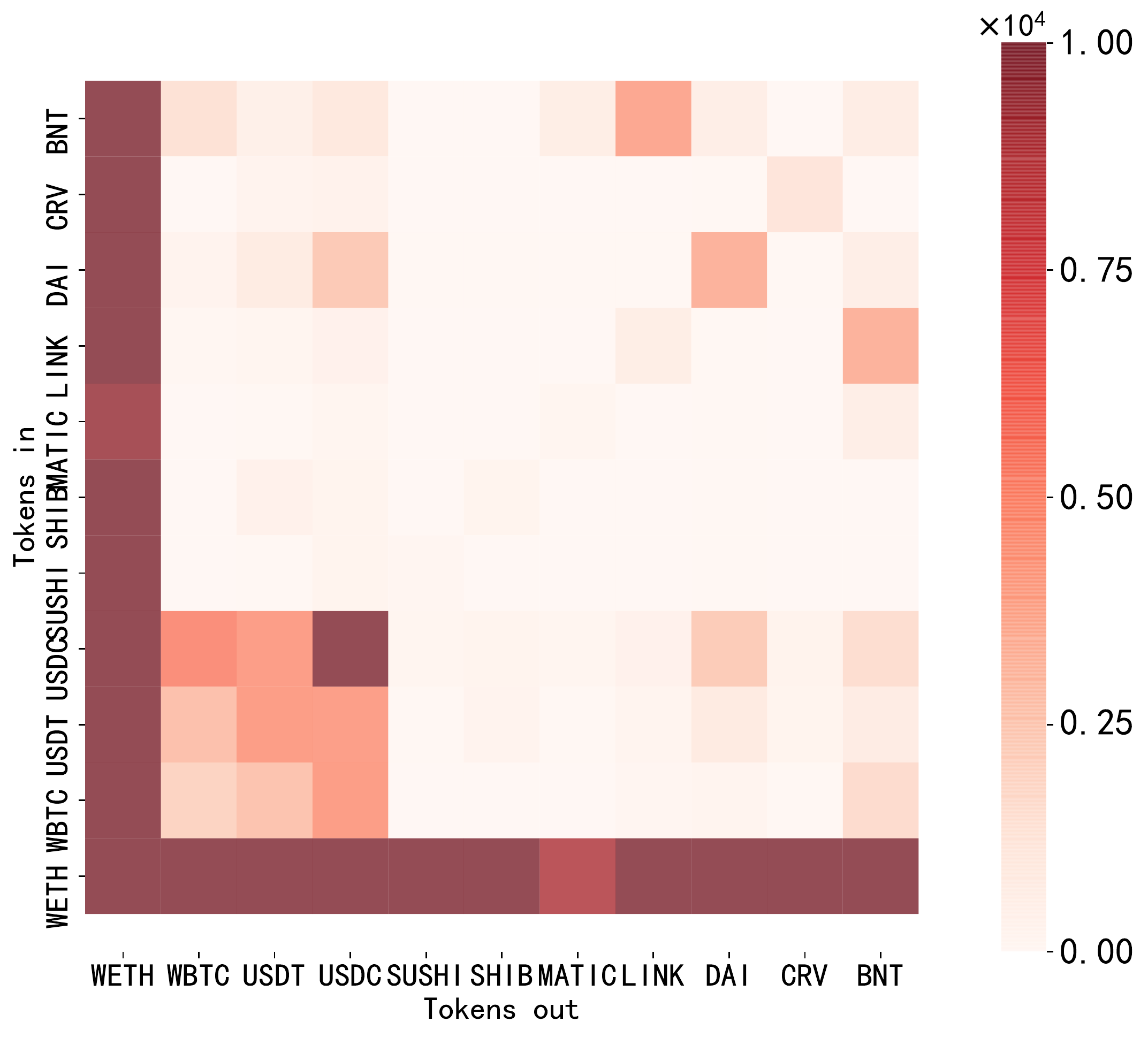}
    \caption{The top token exchange pairs in \pri \txs. Most of the transactions are related to WETH; stable coins such as USDC/USDT are also popular.}
    \label{figs:pairs}
    \end{figure}

    \bheading{The top token exchange pairs in \pri \txs.}
    We measure the most frequently used token exchange pairs of the \pri \txs in ~\figref{figs:pairs}. 
    From the figure, we observe that the exchanges from the stablecoin \textit{USDC} to \textit{WETH} takes the most portion. 
    In addition, the top token exchange pairs usually contain \textit{WETH} and stablecoins.
    It is expected that such tokens are popular on exchanges, since they are less subtle to market changes and more valuable to keep. 
    

    \bheading{The token swap patterns in \pri \txs.}
    To present more details of the complexity and behaviors, we classify the token swap patterns in \pri \txs into the following five types and measure them.
    \begin{packeditemize}
        \item Type I: involving in one token type.
        \item Type II: involving in two types of tokens and two token transfer edges.
        \item Type III: involving in two types of tokens but more than two token transfer edges.
        \item Type IV: involving in three types of tokens.
        \item Type V: involving in more than three types of token.
    \end{packeditemize}

    





    In total, we discover 5,746,633 \pri \txs involved in tokens, which accounts for around 78\% among all the \pri \txs. 
    Specifically, the \pri \txs that only exchange two types of tokens (type II and type III) accounts for about 69\% among all types. Moreover, the percentage of Type V is around 2.2\%, which is the smallest among all types.  
    
    
\bheading{The top DeFi services used by \pri \txs.} 
    We measure the count of \pri \txs against every DeFi.
    Specifically, the top 3 DeFis are \textit{Uniswap V2}~\cite{uniswap-v2}, \textit{Uniswap V3}~\cite{uniswap-v3}, and \textit{SushiSwap}~\cite{sushiswap}, which are related to 558,919, 332,058, and 177,055 \pri \txs respectively.
    Moreover, more than 60\% of the \pri \txs are used for tokens swap in DeFi markets.
    

\subsection{Entities}
We evaluate the entities involved in \pri \txs, to give an overview of the general distribution of different entities. In particular, we measure the top senders, receivers, and miners.  

 




 



%

\bheading{Sender.} 
Senders of \txs are EOAs. 
To give an overview of the senders, we measure the top 10 senders and the count of sent \pri \txs. 
In total, 1,147,181 \pri \txs are sent from the top 10 senders, which is about 15.5\% portion of the total \pri \txs.
The miner address \textit{Ethermine:0xEA674f} is the 1st largest sender creating 581,086 \pri \txs. It mainly uses \pri \txs to transfer money to other addresses for redistributing mining income. 
Another special sender is \textit{Fund:0x0F4ee9}, which is related to investment or venture funds and ranks in the top 8 for sending out 43,995 \pri \txs. 
Such Ethereum addresses that are labeled as \textit{Fund} are usually users who own lots of \textit{ETH} for fund. 
Moreover, the rest of the top senders are all normal EOAs. 


\bheading{Receiver.}
Receivers are the target addresses of \txs, which can either be smart contracts or EOAs. 
We measure the top 10 receivers that are related to 2,275,695 \pri \txs, accounting for about 31\% in total. 
Specifically, all the top 10 receivers are smart contracts. 
From \tabref{tab:top_receivers}, there are 5 MEV Bots involved in 1,037,924 \pri \txs. 
In addition, there are 1,237,771 \pri \txs involving in 4 receivers that are DeFis addresses. 
These DeFi addresses belong to \textit{Uniswap} and \textit{SushiSwap}. 
Morever, the left top receiver is \textit{Tether USDT} stable coin address. 
\Pri \txs calling this address usually are used to transfer the \textit{Tether USDT} to others.


\begin{table}[h]
    \begin{center}
    \setlength{\tabcolsep}{3pt}
    \begin{tabular}{ llr } 
    \hline
    \scriptsize \textbf{Label} & \scriptsize \textbf{Receiver Address} & \scriptsize \textbf{Count} \\
    \hline
    \scriptsize Uniswap & \scriptsize 0x7a250d5630B4cF539739dF2C5dAcb4c659F2488D & \scriptsize 558,919 \\ 
    \scriptsize MEV Bot & \scriptsize 0xa57Bd00134B2850B2a1c55860c9e9ea100fDd6CF & \scriptsize 412,563 \\
    \scriptsize MEV Bot & \scriptsize 0x00000000003b3cc22aF3aE1EAc0440BcEe416B40 & \scriptsize  222,829 \\
   \scriptsize  Uniswap & \scriptsize 0xE592427A0AEce92De3Edee1F18E0157C05861564 & \scriptsize  216,791 \\
    \scriptsize SushiSwap & \scriptsize 0xd9e1cE17f2641f24aE83637ab66a2cca9C378B9F & \scriptsize  177,055 \\
    \scriptsize Token & \scriptsize 0xdAC17F958D2ee523a2206206994597C13D831ec7 & \scriptsize 169,739 \\
    \scriptsize MEV Bot & \scriptsize 0x1d6E8BAC6EA3730825bde4B005ed7B2B39A2932d & \scriptsize 151,668 \\
    \scriptsize MEV Bot & \scriptsize 0x000000000035B5e5ad9019092C665357240f594e & \scriptsize 138,867 \\
   \scriptsize  Uniswap & \scriptsize 0x68b3465833fb72A70ecDF485E0e4C7bD8665Fc45 & \scriptsize 115,267 \\
    \scriptsize MEV Bot & \scriptsize 0x4d246bE90C2f36730bb853aD41d0a189061192d3 & \scriptsize 111,997 \\

    \hline
    \end{tabular}
    \caption{The list of top 10 receivers in \pri \txs and the related count of received \pri \txs.
    }
    \label{tab:top_receivers}
    \end{center}
\end{table} 

\begin{table}[h]
    \begin{center}
    \setlength{\tabcolsep}{3pt}
    \begin{tabular}{ lrrrr } 
    \hline
    \scriptsize \textbf{Miner} & \scriptsize \textbf{\makecell{ Blocks with \\ private \\ \txs}} & \scriptsize \textbf{ \makecell{ Direct \\ payment}} & \scriptsize \textbf{As a sender}  & \scriptsize \textbf{\makecell{ Total \\ private \\ \txs}} \\ 
    \hline
    \scriptsize Ethermine & \scriptsize 316,774 & \scriptsize 482,894 & \scriptsize 581,086   & \scriptsize 1,063,980 \\
    \scriptsize F2Pool & \scriptsize 168,442 & \scriptsize 174,541 & \scriptsize 5,251  & \scriptsize  179,792 \\
    \scriptsize Spark Pool & \scriptsize 143,767 & \scriptsize 310,958 & \scriptsize 969  & \scriptsize 311,927 \\
    \scriptsize Hiveon Pool & \scriptsize 80,937 & \scriptsize 57,088 & \scriptsize 1,872  & \scriptsize 58,960 \\
    \scriptsize Flexpool.io & \scriptsize 46,365 & \scriptsize 93,563 & \scriptsize 484   & \scriptsize 94,047 \\
    \scriptsize MiningPoolHub & \scriptsize 41,701 & \scriptsize 63,691 & \scriptsize 28,010  & \scriptsize 91,701 \\
    \scriptsize Miner:0xb7e...707 & \scriptsize 41,360 & \scriptsize 39,909 & \scriptsize 81  & \scriptsize 39,990 \\
    \scriptsize 2Miners:PPLNS & \scriptsize 36,657 & \scriptsize 39,967 & \scriptsize 1089  & \scriptsize 41,056 \\
    \scriptsize Nanopool & \scriptsize 35,541 & \scriptsize 25,881 & \scriptsize 74  & \scriptsize 25,955  \\
    \scriptsize BeePool & \scriptsize 33,533 & \scriptsize 82,797 & \scriptsize 1,728  & \scriptsize 84,525 \\
    \hline
    \end{tabular}
    \caption{Miners involved in \pri \txs.}
    \label{tab:top_miners}
    \vspace{-10pt}
    \end{center}
\end{table} 

\bheading{Miner.}
In total, 278 miners have been identified and 145 of them have mined blocks with \pri \txs. We list the top 10 miners in table~\tabref{tab:top_miners}, which are sorted by the number of their mined blocks containing \pri \txs. 
Specifically, the table shows the miner, the number of blocks, the number of \pri \txs with direct transfer to the miner, and the number of \pri \txs sent to the miner of the top 10 miners.
Generally, the more blocks the miners have mined, the more \pri \txs with direct transfers will be included and the more profits the miners will earn. In addition, \textit{Ethermine} and \textit{MiningPub} send lots of \pri \txs. We conjecture that these two mining pools are required to redistribute the earned profits frequently or there are many miner nodes in the two mining pools. Moreover, \textit{Sparkpool} (which first introduced \pri 
txs) and \textit{BeePool} stopped mining since September and October of last year, respectively. 


\section{Transaction Cost and Miner Profits} 
\label{sec:money}

In this section, we analyze the transaction cost including used gas and gas price, and measure the miner profits in terms of the distribution, detailed income, and flows. 

\subsection{Transaction Cost}
To present the detailed cost of the transaction fee, we measure the gas used and the gasprice settings from \pri \txs.

    \begin{figure}[t]
    \centering
    \includegraphics[width=1\columnwidth]{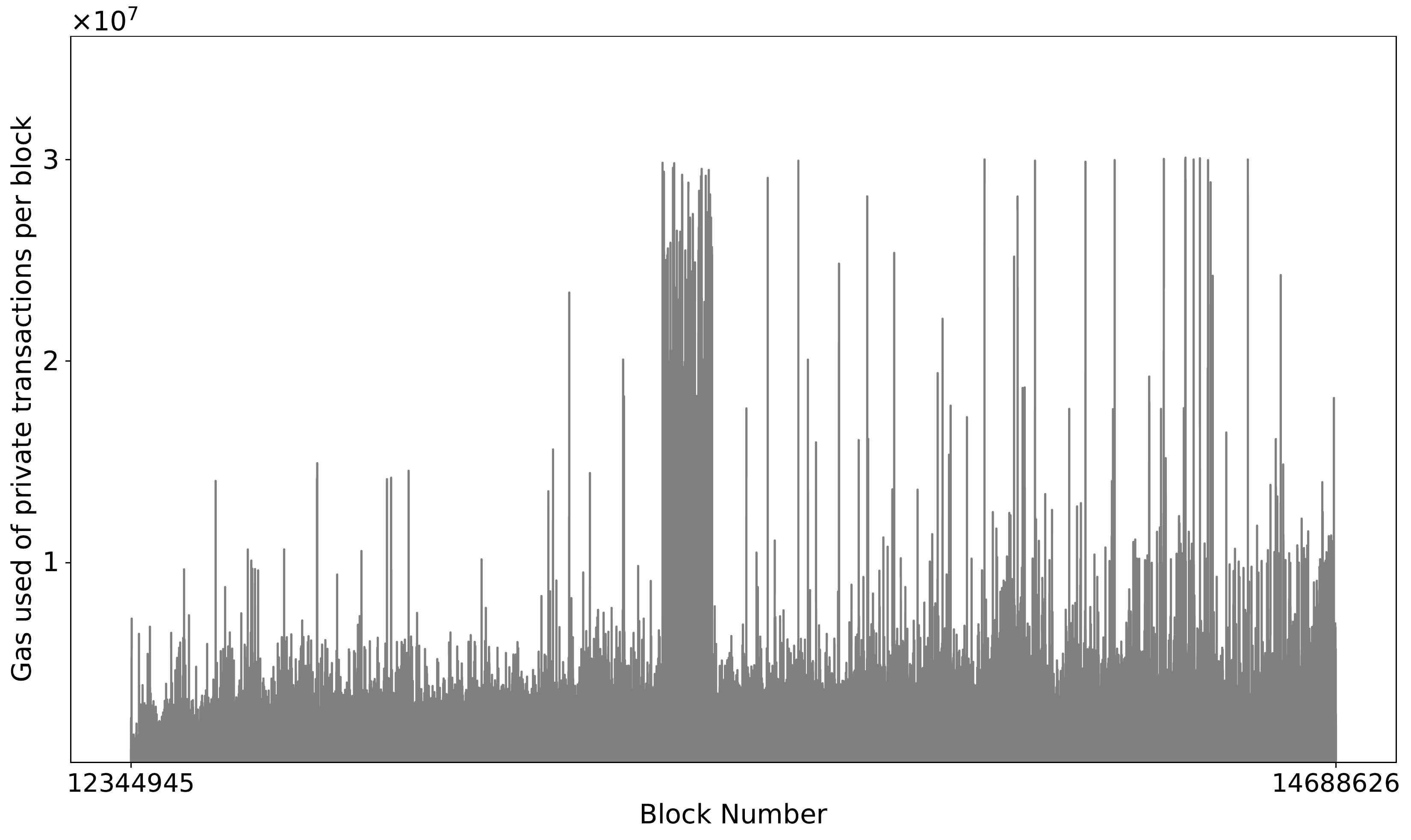}
    \caption{The used gas by \pri \txs per block.}
    \label{figs:usedgas}
    \vspace{-10pt}
    \end{figure}

\bheading{The used gas of \pri \txs.} 
\figref{figs:usedgas} presents the gas used of \pri \txs against per block. 
    The total gas used for \pri \txs in each block is about 737,829 on average, which is much smaller than the average gas used (16,673,757) of normal \txs. 
    Particularly, there are some blocks where the gas used of \pri \txs takes a large portion of the total gas used. It is obvious that \txs in such blocks are mostly \pri \txs.
    For example, the blocks starting from 13,403,072 to 13,497,355 have a high percentage of gas used by \pri \txs, which also relate to the first abnormal period in~\secref{sec:char:dis}.
    For another example, block 14,520,420 has only one \tx and this \tx is a \pri \tx. 
    Meanwhile, there are 367 blocks in total whose used gas percentage by \pri \txs is 100\%. In particular, 361 blocks are mined after the gas limit expansion and 6 blocks are mined before the expansion. 
    
    Moreover, we observe that the expansion of the gas limit helps the growth of \pri \txs. Before the gas limit expansion in block 12,965,000, the used gas by total \txs is about 99\% of the gas limit on average. It shows that the Ethereum network is somehow crowded and \txs are awaiting to be mined. Besides, after the gas limit is increased, miners have the ability to pack more \txs. Accordingly, the number of \pri \txs in blocks can be dynamically increased.  
    

    

\bheading{The gasprice setting in \pri \txs}
    \begin{figure}[t]
    \centering
    \includegraphics[width=1\columnwidth]{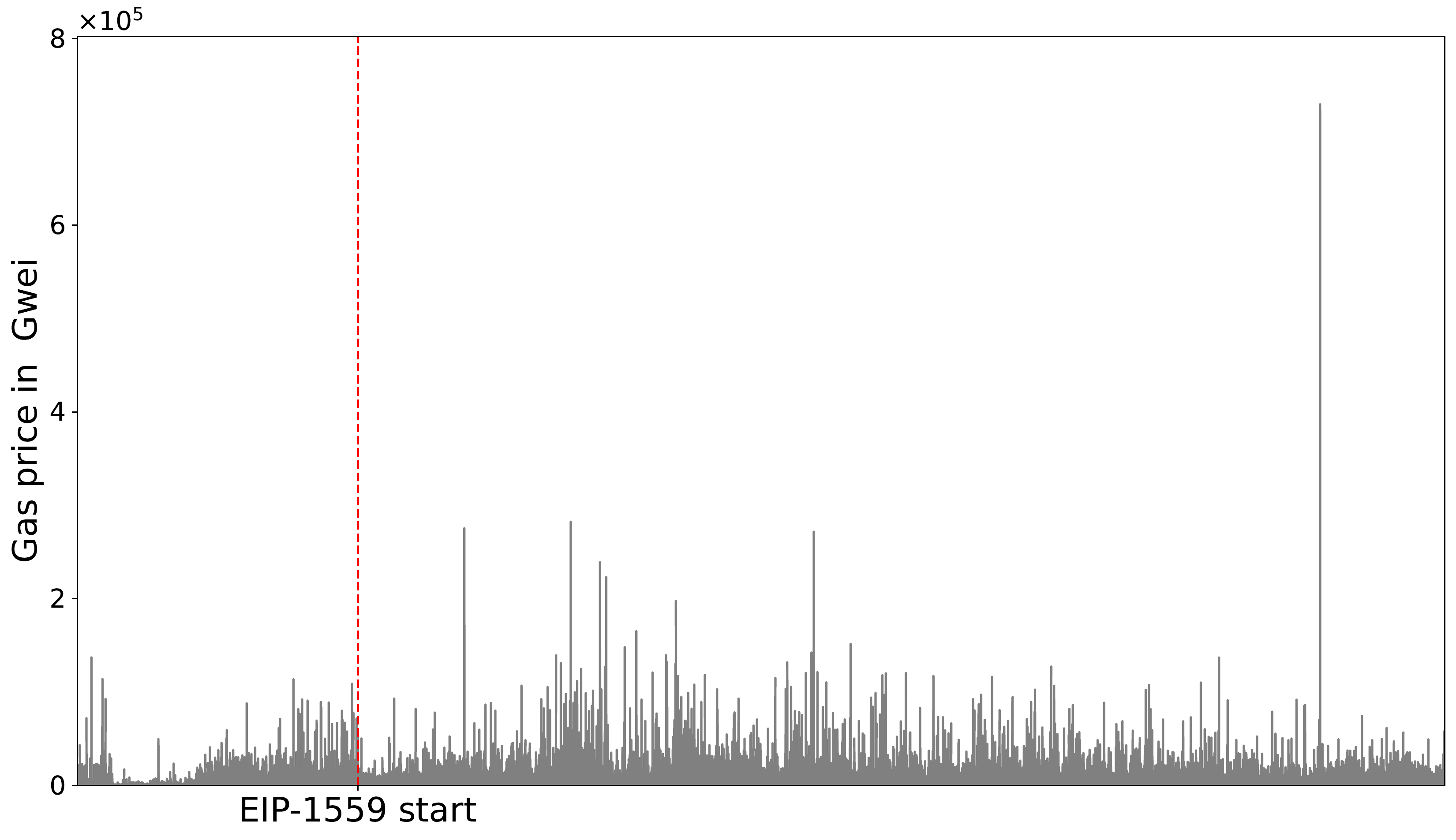}
    \caption{The gasprice of every \pri \tx in chronological order.}
    \label{figs:gasprice}
    \end{figure}
    \begin{figure}[t]
    \centering
    \includegraphics[width=1\columnwidth]{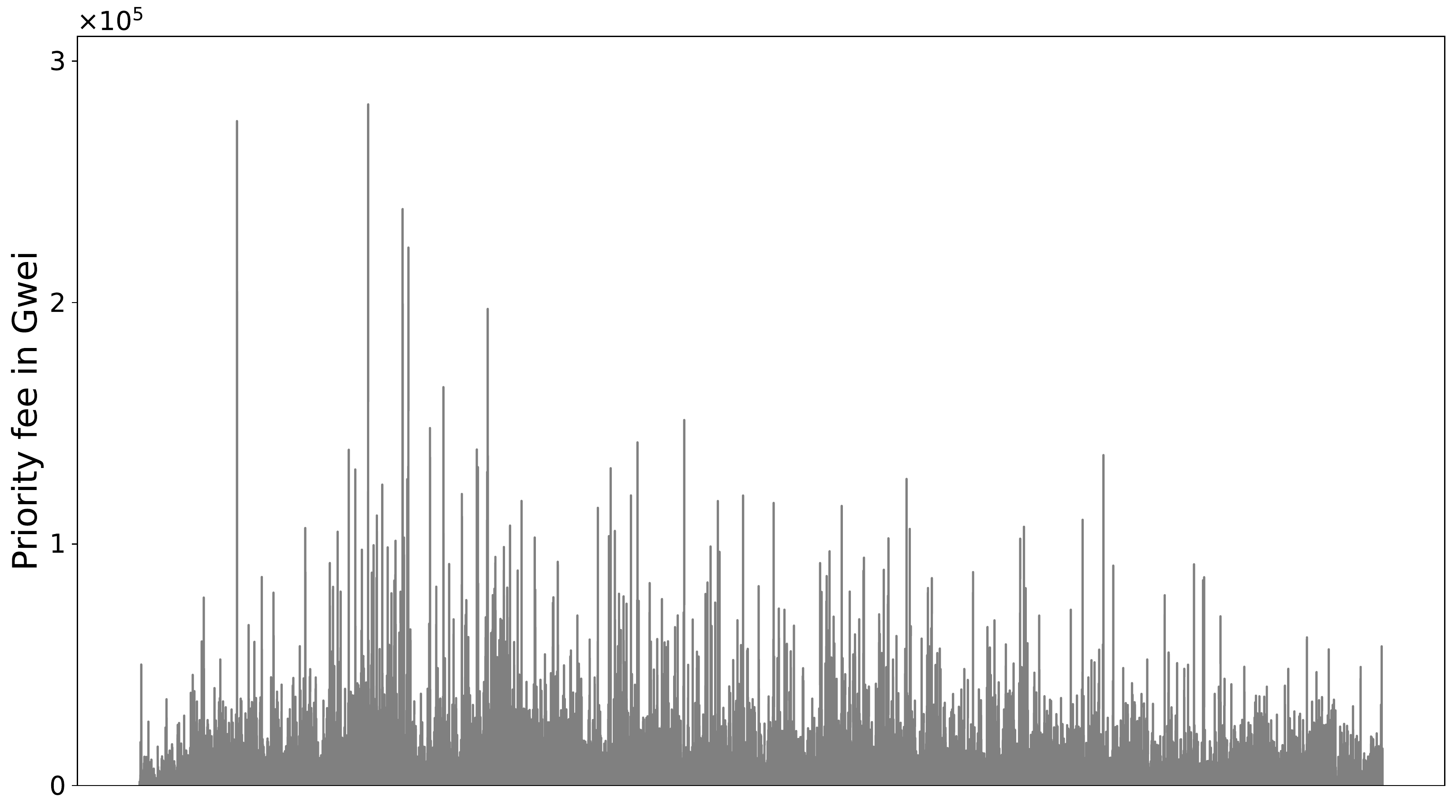}
    \caption{The priority fee of every \pri \tx after EIP-1559 in chronological order.}
    \label{figs:afterpri}
    \vspace{-12pt}
    \end{figure}
    We measure the gas price against every \pri \tx in~\figref{figs:gasprice}. 
    Particularly, we study the influences of EIP-1559 to gas price of \pri \txs. In total, there are 1,528,843 \pri \txs before EIP-1559 and 5,876,992 after EIP-1559.
    
     Before EIP-1559, the average gas price of the \pri \txs is around 93.3 \textit{Gwei}. Moreover, the gas price of 771,979 \pri \txs are set to be 0 accounting for 50\% of the total. Generally, miners will not pack transactions with 0 gas fee into the blocks since they are incentived to max their profits. However, such \pri \txs will pay the earned profits to miners via the direct transfer in the \txs.
    These transactions with 0 gas price are distributed in 318,375 blocks.
    In particular, blocks containing transactions with 0 gas price at least include one direct payment transfer from \pri \txs to miners, which proves that \pri \txs indeed bring profits to miners. Moreover, the average transfer of these \pri \txs is around 0.214 \textit{ETH}. Detailed miner profits will be analyzed later in ~\secref{sec:money:miner}. 
    
    After EIP-1559, the average gas price of all \pri \txs is around 189 \textit{Gwei}. Specifically, we observe that the transaction\footnote{\scriptsize 0xba5a6f7970f5ff78786bb318085f2b6789488827dfec4f8ceed67e289728a7ee} with the highest gas price (729,541.97 \textit{Gwei}) is from a transfer to an EOA address and the total transaction fee is 15.32 \textit{ETH}.
    As required by EIP-1159, the gas price is not allowed to be 0 due to the basefee. \Pri \txs should pay both basefee and priority fee that is the tip for miners, as their gas price. 
    To further measure the transaction fee after EIP-1559, we measure the priority fee against every \pri \tx after EIP-1559.Some \pri \txs did not set the new gas properties such as priority after EIP-1559, and we got 4,105,987 \pri \txs with priority set in~\figref{figs:afterpri}. Among these  \pri \txs, the \pri \tx\footnote{\scriptsize0xc9951933b0aef59f60b5421d39ab89c30ee48be15d80a0db69cdfbe9f72e874a} pays the highest priority fee, only transfers 1 \textit{GWei} from the sender to the sender itself, of which the priority fee is around 282,093 \textit{GWei}. In total, the \pri \tx pays around 6 \textit{ETH} as the transactions. To our knowledge, we doubt that the high-priority fee of this \tx is mistakenly set.
    In addition, there are 1,266,449 \pri \txs whose priority fee is 0,which takes around 22\%  of the total \pri \txs. Moreover, 96.5\% of them are sent to call MEV Bots.
    Transactions with 0 priority fee are distributed in about 615,508 blocks and all of these blocks contain at least one direct payment from \pri \txs to miners. The average transfer amount is around 0.14 \textit{ETH}. 
    As we stated before, while the profits will decrease due to the 0 priority fee, miners are still willing to pack such \pri \txs into the blocks due to the direct transfer serving as a bribe. 
    
     
     \lesson{Some \pri \txs (around 50\% before EIP-1559 and 22\% after) do not use gas fees as rewards to the miners. Instead, MEV profits are distributed to the miners via direct transfers. } 

\subsection{Miner Profits}
\label{sec:money:miner}
In this section, we investigate the mining profits and the flows of the earned profits to assess the impact of \pri \txs on the miners.

\bheading{The distribution of total profits.}
Miners mainly gain benefits from the following three types. 
\begin{packeditemize}
    \item Type I: Block reward. The reward consists of both normal and uncle blocks payouts.
    \item Type II: Profits from \pri \txs. The profits include both transaction fee and the bribe to miners in \pri \txs. Specifically, basefee in \pri \txs will be burnt by the network and deducted from the earned \tx fee for miners.
    \item Type III: Profits from normal \txs. The profits include both transaction fee and the bribe to  miners in normal \txs. The same as Type II, the earned \tx fee in normal \txs will remove basefee.
\end{packeditemize}
\begin{figure}[t]
\centering
\includegraphics[width=.95\columnwidth]{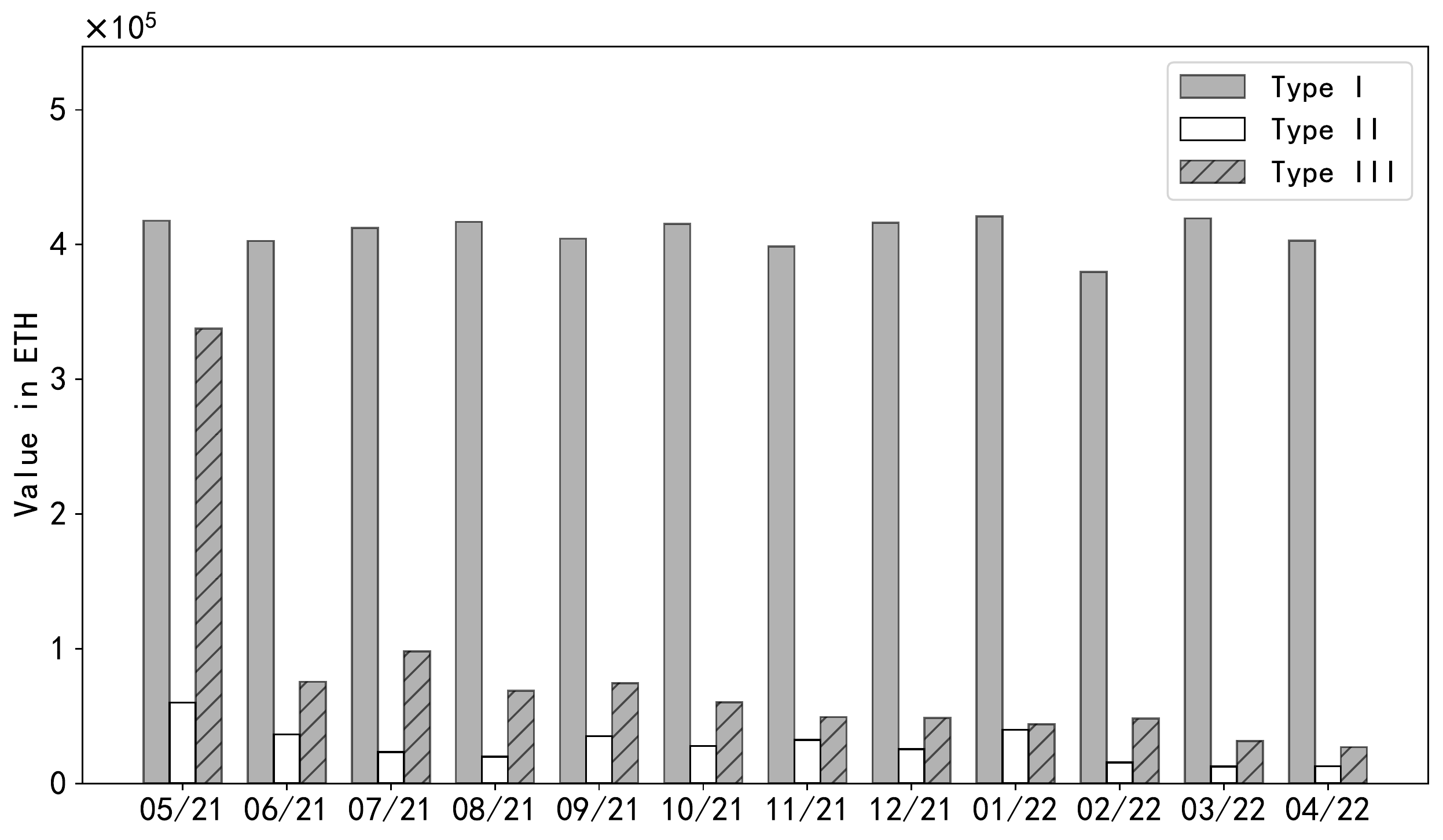}
\caption{The three types of miner profits per month.}
\label{figs:mt}
\end{figure}

We show the profits of each type earned by the miners per month in figure~\figref{figs:mt}. 
From the figure, most of the profits of miners come from the Type I and the amount of the profits, which are almost the same in every single month. 
For Type II, the average monthly profits of the miner is 28,122.8 around \textit{ETH}, accounting for about 5\% of the total miner profit of per month.
Besides, we observe that the profits of Type III is pretty high in May 2021. The reason is that there are about 45 million \txs in that month, which is higher than the average count of 36.5 million. 


\begin{figure}[t]
\centering
\includegraphics[width=1\columnwidth]{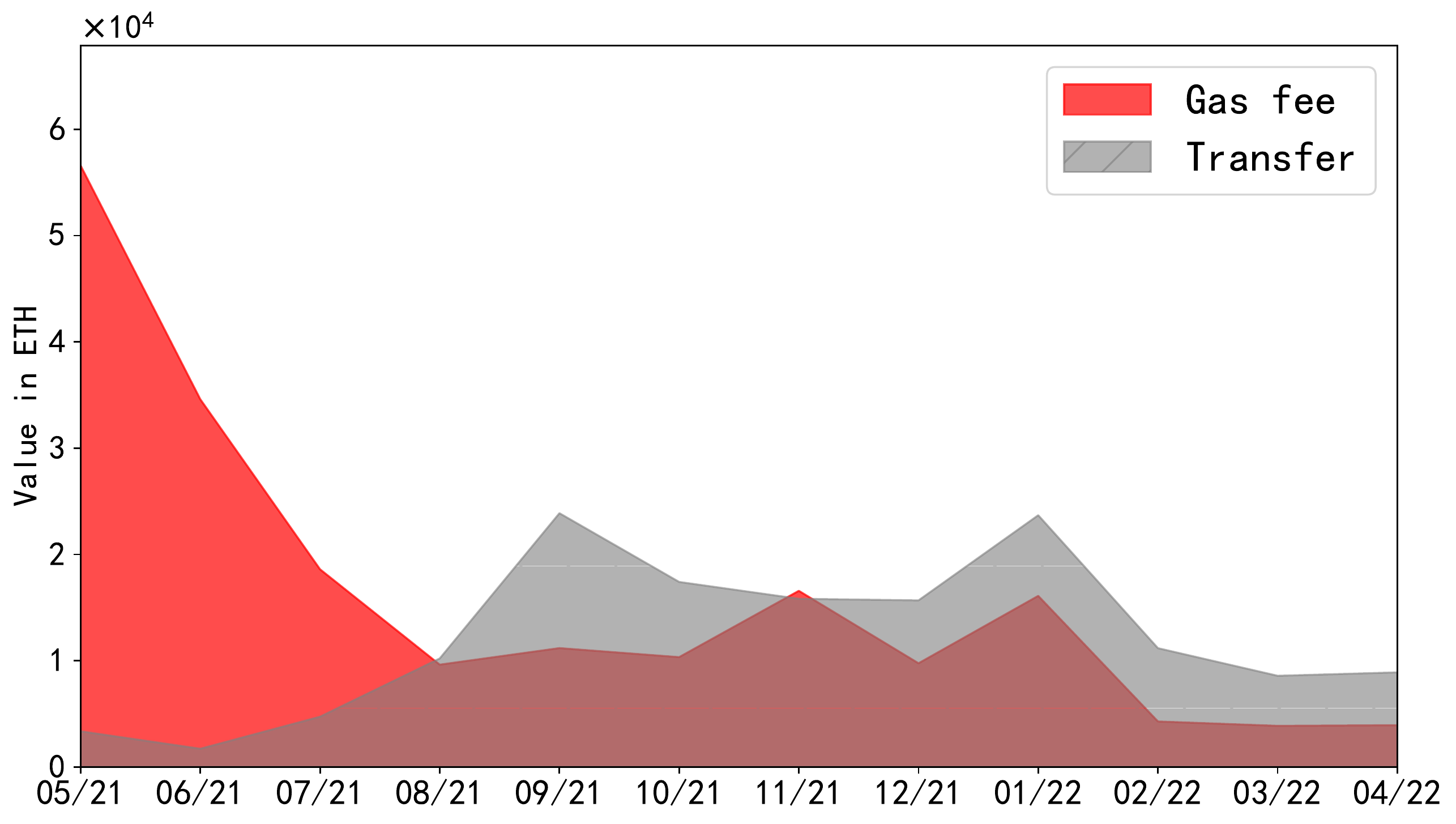}
\caption{The gasfee and transfer in \pri \txs.}
\label{figs:ptp}
\end{figure}

\begin{table}[h]
    \begin{center}
    \begin{tabular}{ lrrr } 
    \hline
    \scriptsize \textbf{Miner} & \scriptsize \textbf{\makecell{ Transfer from \\ private transactions \\ (ETH)}} &  \scriptsize \textbf{\makecell{  Transfer from \\ normal  transactions \\ (ETH)}} & \scriptsize \textbf{\makecell {Total \\ (ETH)}} \\ 
    \hline
    \scriptsize Ethermine  & \scriptsize  59,072.95  & \scriptsize  46,246.64  & \scriptsize  105,319.59 \\ 
    \scriptsize F2Pool Old  & \scriptsize  63,574.24  & \scriptsize  40,400.79  & \scriptsize  103,975.03 \\ 
    \scriptsize Spark Pool  & \scriptsize  20,578.58  & \scriptsize  164.34  & \scriptsize  20,742.92 \\ 
    \scriptsize BeePool  & \scriptsize  5,677.03  & \scriptsize  40.81  & \scriptsize  5,717.84 \\ 
    \scriptsize Flexpool.io  & \scriptsize  5,002.37  & \scriptsize  73.42  & \scriptsize  5,075.79 \\ 
    \scriptsize Hiveon Pool  & \scriptsize  4,631.24  & \scriptsize  84.45  & \scriptsize  4,715.69 \\ 
    \scriptsize MiningPoolHub  & \scriptsize  3,057.21  & \scriptsize  277.08  & \scriptsize  3,334.29 \\ 
    \scriptsize 2Miners:PPLNS  & \scriptsize  2,443.37  & \scriptsize  127.89  & \scriptsize  2,571.26 \\ 
    \scriptsize Miner:0xb7e390  & \scriptsize  2,210.07  & \scriptsize  21.79  & \scriptsize  2,231.86 \\ 
    \scriptsize Nanopool  & \scriptsize  1,708.91  & \scriptsize  57.67  & \scriptsize  1,766.58 \\ 
    \hline

    \end{tabular}
    \caption{The transfer profits in both \pri \txs and normal \txs of top 10 miners.}
    \label{tab:miner_transfer}
    \vspace{-10pt}
    \end{center}
\end{table}

\bheading{The distribution of profits from \pri \txs.}
In particular, we present the transaction fee and direct transfer to miners of \pri \txs against per month in~\figref{figs:ptp}. In total, around 23\% of all \pri \txs transfers money to miners, and the total transferred amount is around 195,381 \textit{ETH}, which accounts for around 58\% among the total revenue of 337,474 \textit{ETH} form \pri \txs. In the figure, the profits from the direct transfer exceed the transaction fee in some months. The reason is that some \pri \txs set low priority fee and pay the miners with the direct transfer. 
Additionally, we list the top 10 miners sorted by the income from direct transfers of \pri \txs in~\tabref{tab:miner_transfer}. Specifically, the income of the top 3 miners accounts for around 90\% of the total. From the table, we notice that most of the transfers come from the \pri \txs, which demonstrates that \pri \txs bring in a lot of money for miners.


\bheading{The flow of earned profits from miners.}
We observe that miners frequently transfer a certain amount of \textit{ETH} to others. For example, some miners usually send 0.02 \textit{ETH} to particular addresses. These \txs are used by miners to redistribute the income to the mining nodes~\footnote{A miner normally groups multiple mining nodes into a pool to combine the computation powers, and then splits the rewards to these mining nodes~\cite{mining-node}.} that perform the mining work. The money is to reward the computation power of these nodes. 
Specifically, the profit redistribution \txs is always \pri \txs for two reasons. The first reason is that miners are capable of sending \txs by themselves to avoid the long waiting time in blockchains. The second reason is that miners can save money since they do not have to pay the priority fee and earn the basefee they pay. 

\begin{table}[h]
    \begin{center}
    \begin{tabular}{ lrrrrr } 
    \hline
     \scriptsize \textbf{Address}  &  \scriptsize \textbf{\makecell{Txs \\ Inbound}} &	 \scriptsize \textbf{\makecell{Total \\ Miner \\ Inbound}}  &	 \scriptsize \textbf{\makecell{ \Pri  txs  \\ Inbound}} &	 \scriptsize \textbf{\makecell{Txs \\ Outbound}} &	 \scriptsize \textbf{\makecell{Exchange \\ Outbound}}  \\
    \hline
    \scriptsize  0x909bFe  & \scriptsize 509 &\scriptsize  509 &\scriptsize 58 &\scriptsize  152 &\scriptsize  152 \\
   \scriptsize  0x18690F & \scriptsize 492 & \scriptsize 468 &\scriptsize 67 &\scriptsize  191 & \scriptsize 185  \\
   \scriptsize  0xE426ec &\scriptsize  476 & \scriptsize 476 &\scriptsize 64 &\scriptsize  49 &\scriptsize  49  \\
   \scriptsize  0xa4C916 &\scriptsize 327 &\scriptsize  312 &\scriptsize 72 & \scriptsize 23 & \scriptsize 23  \\
   \scriptsize  0xbaAa28 &\scriptsize  341 &\scriptsize  341 &\scriptsize 52 &\scriptsize  4 &\scriptsize  0 \\
    \hline
    \end{tabular}
    \caption{The address prefix, Transaction Inbound (the number of \txs sent to the address), Ethermine Inbound (the number of \txs Ethermine sends to the address), Transaction Outbound (the number of \txs this address sends out), and Exchange Outbound (the number of \txs sent from the address for token exchanges) of the top 5 addresses, which are sorted by the count of Ethermine Inbound.}
    \label{tab:miner_flow}
    \end{center}
\end{table} 

To study the detailed flows of the profits of the miner, we performed a case study on the top miner \textit{Ethermine} who sends out 581,086 \pri \txs. 
We list the top five addresses receiving the largest number of \pri \txs from \textit{Ethermine} in~\tabref{tab:miner_flow}. These five addresses are all EOA addresses. From the table, we observe that most of such addresses are used as the wallets of miner nodes, which normally receive the profits from \textit{Ethermine} and then withdraw money. For example, the address~\footnote{0x909bfe97a6f77d2bec31b4c19e8c2810136abb90} receives 509 \txs and all are transfers from miners, in which 58 \txs are \pri \txs. Additionally, the address sends out 152 \txs to withdraw money from their personal wallets via DeFi platforms (e.g. \textit{Binance}). The behaviors are pretty similar for the other four addresses. With the investigation, we have a better understanding of the money flows of the miner profits.  
\looseness=-1

\lesson{Miners take a major share of the profits of \pri \txs. Miners, in turn, re-distribute the profits to mining nodes via \pri \txs, rewarding their contribution in the computing power.} 

\section{Security-related Issues}
\label{sec:security}

In this section, we present the measured results to quantify the related security issues, including MEV, real-world DeFi attacks, consensus security, and \pri \txs leakage.
Moreover, we perform detection on \pri \txs from our one-year dataset in Appendix to study arbitrage, which is one popular type of MEVs. 
 
\subsection{MEV}

    \bheading{The earned profits by MEV Bots.}
    We have identified 143 MEV Bots participating in the total \pri \txs and they are involved in 2,116,587 \pri \txs. 
    Among these \pri \txs, 766,640 of them are identified to gain profits. 
    \tabref{tab:bot_profit_range} shows the \pri \txs distribution of the earned profit. Specifically, 2.6\% calling MEV bots gain more than 10 \textit{ETH} as profits, while 92.8\% earn less than 1 \textit{ETH}.
    Furthermore, we find the average \tx cost of these \pri \txs is around 0.005 \textit{ETH} and 60\% of them share the miner with an average transfer of 0.2 \textit{ETH}.
    Moreover, we identify about 3\% of these \pri \txs ending up with negative pure profits after deducting \tx cost and miner profit share from the earned profits.
 
    
   \begin{table}[h]
    \begin{center}
    \begin{tabular}{ lrrrrr } 
    \hline
    \scriptsize \textbf{{{Profit Range (in ETH)}}} & \scriptsize  [0,1)    & \scriptsize   [1,10)   & \scriptsize   [10,50)   & \scriptsize  [50,100)    & \scriptsize  $\geq{100}$ \\
    \scriptsize \textbf{Count} & \scriptsize 711,441   & \scriptsize   35,265   & \scriptsize  7,794   & \scriptsize   2,830   & \scriptsize     9,241 \\
    \scriptsize \textbf{Percentage} & \scriptsize 92.8\%    & \scriptsize   4.6\%    & \scriptsize   1.02\%   & \scriptsize    0.37\%   & \scriptsize     1.21\% \\
    \hline
    \end{tabular}
    \caption{The total profits distribution of profitable MEV related \pri \txs.}
    \label{tab:bot_profit_range}
    \end{center}
\end{table} 
    
    \bheading{Failed \txs.}
    Generally, users need to pay the transaction fee even though their \txs fail, of which the fee is decided by both gas price and used gas. However, before EIP-1559, the gas price can be set to 0 and users can pay nothing for \txs, regardless of whether they fail or not. \Txs with 0 gas price are not usually mined to blockchains since they give miners no profits in terms of the \tx fee. But with the appearance of \pri \tx, users pay or bribe miners by directly transferring money to them. Thus, miners are willing to mine such \txs. 
    Specifically, we observe 94 \pri \txs fail before EIP-1559 within our dataset and 58 of them have 0 gas price. 
    Although EIP-1559 forces the \txs to pay at least the basefee, \pri \txs users can set the priority fee to 0, to reduce the potential loss if their \txs fail. We observe 19,762 failed \pri \txs and 577 of them have 0 priority fee after EIP-1559. 

    
    
    \lesson{Allowing zero gas price before EIP-1559 in \pri \txs enables attacks with no economic loss, whereas the basefee introduced by EIP-1559 increases costs for failed attacks.}

\subsection{Attack Case Studies}
We investigate three DeFi related attacks that are exploited via \pri \txs. Specifically, we perform case studies on \textit{PolkaBridge}~\cite{polkabridge}, \textit{LI.FI}~\cite{lifi}, and \textit{Multichain}~\cite{mutlichain} attacks. Moreover, we measure the top 30 \pri \txs according to the money transferred to the miners and the top 10 blocks sorted by the amount of profits from the miners.

\bheading{PolkaBridge Attack.}
\textit{PolkaBridge} is a cross-chain DeFi and the first to decentralized bridge between \textit{Polkadot} platform and other blockchains. \textit{Polkadot}~\cite{wood2016polkadot} is a blockchain protocol connecting multiple blockchains into one unified network. \textit{PolkaBridge} mainly is used to support the token exchanges between multiple blockchains, which is also what the bridge aims to. 
\textit{PolkaBridge} was attacked due to a smart contract logic error at Nov-22-2021 07:44:02 AM +UTC.  
In particular, an attacker can withdraw tokens that did not belong to them by simply calling the related function. Moreover, The attacker stole around \$632K worth of assets via the transaction\footnote{\scriptsize 0x36d2f3aaf7d160ea9bd072692555e6d3ff9b76139c8dfa83a475b23cc39cf8e6}. 

As labeled by Etherscan, this transaction is a \pri \tx in Ethereum blockchain and mined in block 13,663,243. Besides, the \tx was placed in the first place inside the block by miner \textit{Poolin 3}\footnote{\scriptsize 0x2a20380dca5bc24d052acfbf79ba23e988ad0050}. Specifically, the attacker set the priority fee at 0 and transferred 60.1012422 \textit{ETH} to the miner as a bribe. This attack case shows that attackers can utilize \pri \txs to collude with miners and attack the exploited platforms. As far as we know, there is no related report or paper that analyzes the bribe to the miner. We are the first to investigate the relationship of attack, miner, and \pri \tx. 

\bheading{LI.Fi Attack.}
We studied another attack on DeFi platform \textit{LI.FI} supporting cross-chain bridging, swapping, and messaging. The services provided by \textit{LI.FI} are similar to the services from \textit{PolkaBridge}, which are for token exchanges between multiple blockchains. According to the report~\cite{lifi-attack}, \textit{LI.FI} was attacked due to the unchecked external call vulnerability in smart contracts. The attack happened on Mar-20-2022 02:51:44 AM +UTC via the transaction\footnote{\scriptsize 0x4b4143cbe7f5475029cf23d6dcbb56856366d91794426f2e33819b9b1aac4e96}, which caused a huge financial loss around \$596K. Moreover, the transaction was mined in block 14,420,687 and the miner was \textit{MiningPoolHub}. As expected, the \tx was placed in the first place inside the block.

Although the attacking \tx is label as \pri \tx by Etherscan, the attack did not share the earned profits with the miners. After investigating the token flow of the attacking \tx, we did not find any direct transfer to the miner and the priority fee was set to be 2.2 \textit{GWei}. We suspect that the fee of launching a \pri \tx was paid off-chain by the attacker. 
From these two attacks, we can conclude that attacking \txs have the urges to not be monitored by others and placed in the front of the blocks, in case that attack competitors take the opportunity away.

\bheading{Multichain Attack.}
\textit{Multichain} is also a DeFi platform for cross-chain services. It was attacked due to the unauthorized permission, according to the report~\cite{mutlichain-attack}. Specifically, the vulnerable smart contract function did not validate the token input parameter and the call being successful or not. 
Moreover, \textit{Multichain} was attacked by at least three attacker groups;
there were multiple attacking \txs and the total stolen amount was estimated to be around \$3 million. After investigation, lots of them are labeled as \pri \txs. 

In particular, we studied the attacking \tx\footnote{\scriptsize 0xe50ed602bd916fc304d53c4fed236698b71691a95774ff0aeeb74b699c6227f7} with the highest financial loss, accounting for 308 \textit{ETH} (around \$950K). This \tx behaves as the attacking \tx in \textit{PolkaBridge} attack. Similarly, it set the priority fee to be 0 and shared 274.686613834949640784 \textit{ETH} to the miner. Moreover, the \textit{from} address\footnote{\scriptsize 0xfa2731d0bede684993ab1109db7ecf5bf33e8051} is labeled as both \texttt{Multichain Exploiter 4} (the fourth found attacker exploiting this attack) and \texttt{Whitehat} (who aims to rescue the attacks). 
The reason is that at first the address hacked the smart contracts for attacks and then agreed to return part of the earned profits after negotiation~\cite{multichain-return}. In addition, \tx is placed at the very beginning of the block 14,037,237 by the miner ~\textit{Ethermine}\footnote{\scriptsize 0xea674fdde714fd979de3edf0f56aa9716b898ec8}.
For another example, attacking \tx\footnote{\scriptsize 0x6e26e226d81563a96e6aa9423b1301ce798242b5b4387ad7b9ac5b66966c2a44} (labeled as a \pri \tx) was launched by another attacker address\footnote{\scriptsize 0xd37448ad7949c4ad8eba5aad1a0afdd3199971d8}, which is labeled as \texttt{Multichain Exploiter 7}. The patterns between the two attacking \txs are exactly the same.
From these example \txs, we observe that \pri \txs indeed help the attackers to take the priority for attacks.  \looseness=-1


\lesson{Real world attacks have used \pri \txs to seize the fleeting attack opportunities. In some cases, the miners take a major share of the profits, making them unwitting accomplices in these attacks.}

\begin{figure*}[tb]
\centering
\begin{subfigure}[b]{0.49\textwidth}
\includegraphics[width=\textwidth]{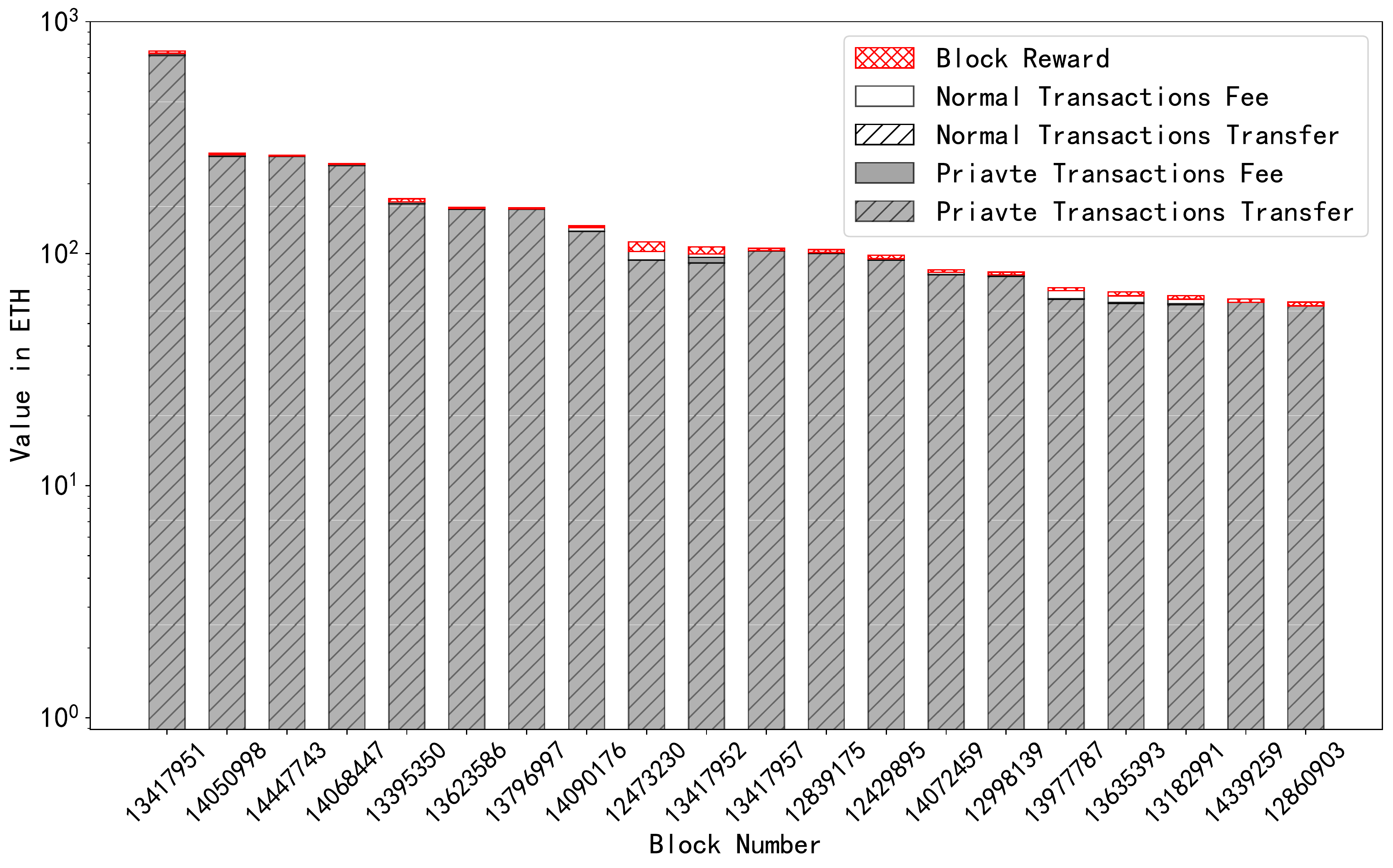}
\caption{Top 20 blocks with private \txs.}
\label{figs:top_blocks}
\end{subfigure}
\hfill
\begin{subfigure}[b]{0.49\textwidth}
\includegraphics[width=\textwidth]{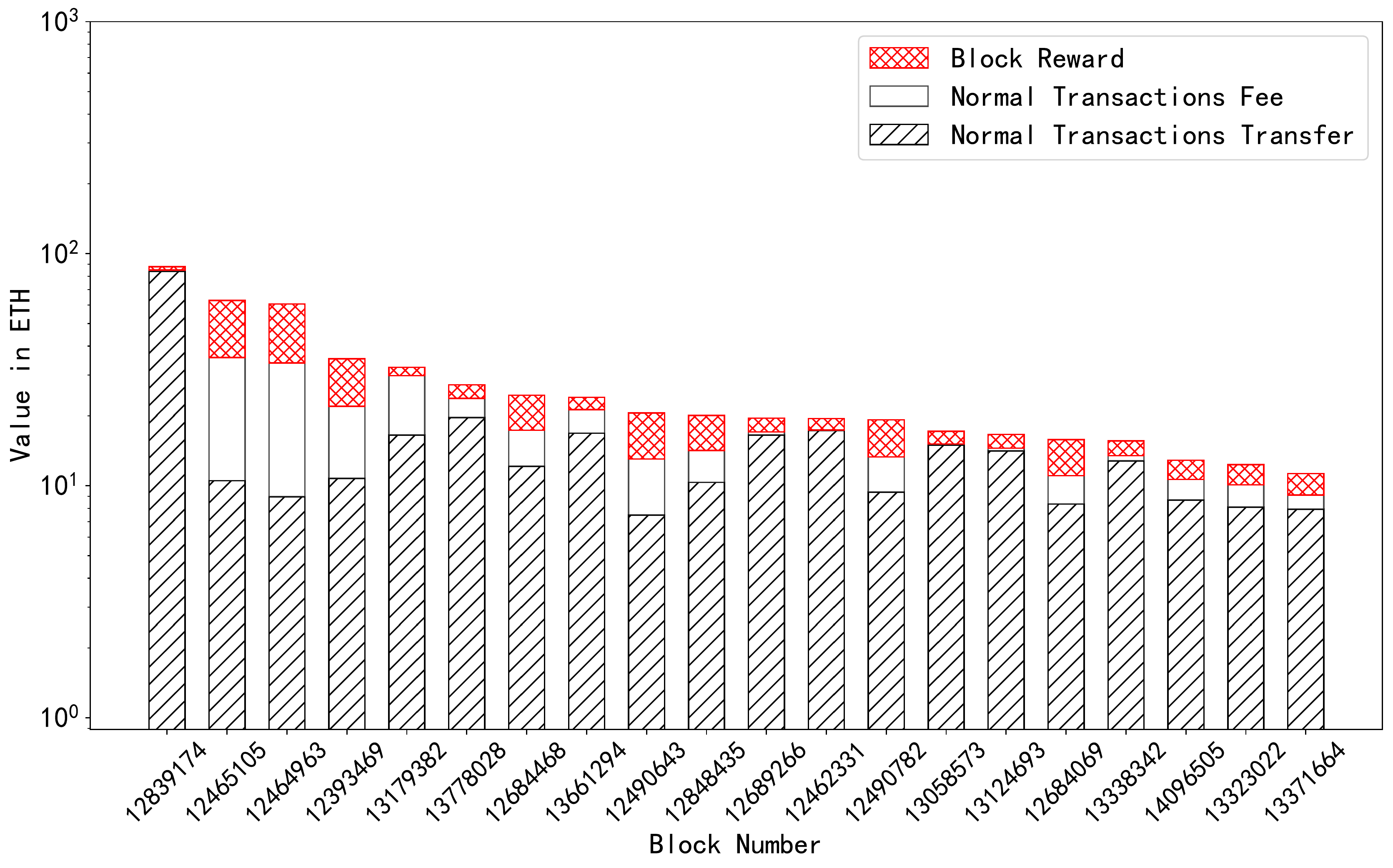}
\caption{Top 20 blocks with normal \txs only.}
\label{figs:top_normal_blocks}
\end{subfigure}
\caption{Comparison: Top 20 blocks containing \pri \txs, and top 20 blocks with normal \txs only.}
\label{fig:top-20-block}
\end{figure*}

\subsection{Consensus Security}
We measure the most profitable \pri \txs from the user side and blocks for the miner side, and find that \pri \txs have threats on the consensus security. Specifically, the huge profits brought by \pri \txs increase the risks of undercutting attacks.  

\bheading{The top 50 \pri \txs.} 
After manually checking the top 50 \pri \txs sorted by the transfer amount to miners, 49 of them are related to MEV and 1 is related to real-world attacks, which is the \textit{Multichain} attack we mentioned above. In addition, MEV related \txs are used for arbitrage (34 among 50 \pri \txs) or liquidation (15 among 50 \pri \txs). 
For arbitrage, these \pri \txs usually interact with \textit{Uniswap}, \textit{SushiSwap}, and \textit{Curve}. For liquidation, \pri \txs are usually sent to \textit{Aave}, \textit{dYdX}, and \textit{Compound}.
Specifically, the most profitable \tx\footnote{\scriptsize 0x3b5fc9f804e026a23f1474218bc9d15e69a4fded405c8ffda0beb028e9d61c53} is used for arbitrage. It earns 729.6836279 \textit{ETH} and shares 706.3337518 \textit{ETH} with the miner. Besides, the \pri \tx costs 0.03407946836279 \textit{ETH} as the \tx fee, while the priority fee is set to 0. 

\bheading{The top 20 blocks.}  
To study the influences of \pri \txs, we present the top 20 blocks that earn the most profits in~\figref{figs:top_blocks}. 
From the figure, we observe that most of the profits in the top 20 blocks come from the direct transfers from \pri \txs to miners, which takes around 90\% among the total profits in blocks.  
In particular, the top block is 13,417,951 and earns around 747 \textit{ETH}, in which the transfer from \pri \txs is about 714 \textit{ETH}. The most profitable \pri \tx we mentioned above comes from this block and transfers  706.3337518 \textit{ETH} to the miner. We can learn that other profits types only take a very small portion.
For comparison, we show the top 20 blocks without \pri \txs in~\figref{figs:top_normal_blocks}, sorted by the earned profits from miners. The most profitable block 12,839,174 earns around 88 \textit{ETH}, which is far less than the profits from top blocks with \pri \txs in~\figref{figs:top_blocks}.
Moreover, the results shown in~\figref{figs:top_normal_blocks} are comparable with those presented by Daian~\etal~\cite{flashboy}: the most profitable block (block 7,029,147) resulted in around 100~\textit{ETH} profit as well.
\looseness=-1

\bheading{Undercutting attacks.}
An undercutting attack happens when the profits from \txs exceed the block rewards, in which miners deliberately fork the existing blockchain and leave these profitable transactions unclaimed, in order to attract other miners to create new blocks on the undercutting blockchain. To successfully launch the undercutting attacks, attackers need to give enough money incentives to the miners since the main goal of a rational miner is to maximize its profits.
Previously, Carlsten~\etal~\cite{carlsten2016instability} consider the \tx fees as the only factor that causes undercutting attacks; 
In~\cite{flashboy}, it also takes Ordering Optimization (OO) fees (including MEV) into account. 
According to Gong~\etal~\cite{gong2020towards}, 
before private \txs were introduced,
the money deliberately left by attackers may not be enough to 
incentivize the miners to launch the undercutting attack. 
For instance, the \txs that are left for attracting miners might not be able to fit into the blocks.
However, from the above experiments measuring the top 20 blocks in 
both private \txs and normal \txs (\figref{figs:top_blocks} 
and \figref{figs:top_normal_blocks}), 
\pri \txs greatly increase the incentives for undercutting attacks; 
miners can easily find the most profitable \pri \txs 
and place them into the new block by squeezing out the less profitable ones. 
Therefore, we find that \pri \txs can be a potential driving factor  
that leads to undercutting attacks 
since the block rewards are far lower than other profits, 
especially the direct transfers from \pri \txs. 

\lesson{Private \txs exacerbate threats to consensus protocols. The huge profits brought by \pri \txs might provide the miners additional incentives to launch undercutting attacks.}

\subsection{Leakage of \Pri \Txs}
\label{sec:security:leak}
We notice that there are some inaccuracies of the \pri \txs label from Etherscan. For example, the \tx\footnote{\scriptsize 0x8d0c16210335a9ee8815d7b0dba22134f9dc722047e8f2d67399184cd92c420f} is labeled as a \pri \tx in Etherscan, but found it in our local mempool. 

\begin{table}[t]
    \begin{center}
    \scriptsize
    \setlength{\tabcolsep}{5pt}
    \begin{tabular}{ lrrrrr } 
    \hline
 \scriptsize \textbf{Date} & \scriptsize \textbf{\makecell{Node 1 \\ View}} &  \scriptsize \textbf{\makecell{Node 2 \\ View}} &  \scriptsize \textbf{\makecell{Two Nodes \\ Combined}} &   \scriptsize \textbf{\makecell{Labeled by \\ Etherscan}} &   \scriptsize \textbf{\makecell{Found in \\ mempool}} \\ 
\hline
 May 22 &   726,225 &   935,288 &   1,015,771 &   32,831 &   1,272 (3.9\%)  \\ 
 May 23 &   735,962 &   1,025,886 &   1,060,942  &  24,162 &   1,350 (5.6\%) \\ 
 May 24 &   766,517 &   1,035,457  &   1,059,803 &   33,787 &   1,381 (4.1\%) \\ 
 May 25 &   739,896 &   971,091 &   1,010,729 &   30,708 &   1,686 (5.5\%) \\ 
 May 26 &   776,464 &   1,000,375 &   1,011,134 &   21,669 &   1,109 (5.1\%) \\ 
 May 27 &   727,589 &   1,010,602 &   10,20,657 &   31,642 &   1,346 (4.3\%) \\ 
 May 28 &   760,488 &   987,575 &   1,027,682 &   26,772 &   1,047 (3.9\%) \\ 
 May 29 &   752,500 &   951,380 &   1,005,138 &   31,158 &   1,072 (3.4\%) \\ 
 May 30 &   735,069 &   971,857 &   1,041,587 &   32,831 &   1,319 (4.0\%) \\ 
    \hline
    \end{tabular}
    \caption{The \pri \txs in the mempool of two nodes and comparison with Etherscan.}
    \label{tab:mempool}
    \vspace{-8pt}
    \end{center}
\end{table}

To study the inaccuracies, we deploy two modified full nodes in two continents to monitor the \txs coming to the local mempool from the nine-day dataset. Then we compare the \txs collected from the mempool of two nodes with the \pri \txs from Etherscan and present our results in~\tabref{tab:mempool}.
From the table, we observe a 4.3\% leakage in 9 days with the two nodes.
We suspect that some \pri \txs are leaked due to the uncle block, block reorganization, the orphan \tx, the loss with the low gasprice, and so on.
For example, when a \pri \tx is successfully packed into the block but is then considered as an uncle block, \pri \tx will be released to the local mempool of every node. 

The leakage of \pri \txs will harm the profits of users. For example, if the user of a \tx exchanges a large amount of tokens and tries to avoid being monitored by others, he/she will submit a \pri \tx. However, assuming the \pri \tx is leaked and frontran by MEV Bots, it will cause a huge financial loss to the user. Similarly, the leakage can cause the failure of MEV opportunities. For instance, if the MEV related \pri \txs is leaked into the mempool and placed in the lower places of the mined blocks, the competitors are likely to take advantage of the MEV opportunity before them.  
\lesson{We observe \pri \tx leakage from our two local Ethereum nodes. Therefore, \pri \txs are not guaranteed to be absent from the public mempool; they may still be vulnerable to attacks.}


\section{Discussion}
\label{sec:discuss}

\bheading{Limitations.}
In this paper, we use the labels from Etherscan as our ground truth,
similar to previous works~\cite{kumar2020detecting,valadares2021identifying,agarwal2021vulnerability}.
Etherscan is  trusted by the community that it has a clear and comprehensive view of the labels, since Etherscan has reliable data sources.
For \pri \tx labels, Etherscan deploys Ethereum nodes deployed over the world to capture \pri \txs. 
Besides, they fetch the data from \pri \tx service providers (e.g., Flashbots~\cite{flashbot}) to label these \txs. 
For instance, there is a label called Flashbots in Etherscan Label World, which provides all the \pri \txs coming from Flashbots.
However, Etherscan might still have mis-labeled data on \pri \txs.
To enhance the authenticity of the \pri \txs labels, 
we may consult other third-party mempool services (e.g., BLOXROUTE~\cite{blockxroute}) 
and cross-check their labels with Etherscan's labels. 
We leave this as our future work.

\bheading{Future extensions.}
As stated in~\secref{sec:security:leak}, there is \pri \tx leakage at around 4.3\% percentage. The leaked \pri \txs against their intentions and may harm the profits of their users. We can analyze the purposes and usages of these \pri \txs and study the potential risks. Besides, there are 299 \pri \txs are leaked not due to the uncle blocks. It might bring some interesting insights to measure how they are leaked. 
Moreover, in this paper,
we perform analysis on arbitrage in private \txs in the Appendix.
It would be interesting to measure other attacks such as 
sandwich attacks and flashloans in the private \tx pool.
We leave these to the future work.

\bheading{Ethereum 2.0.}
Ethereum will move from PoW to PoS and replace miners with validators. Users can become validators as long as they stake 32 \textit{ETH} as collateral that encourages them to behave well. Specifically, validators are either chosen to create blocks or responsible for validating blocks. 
With the upgrade of the consensus layer, \pri \txs will not disappear. Instead, MEV opportunities will attract more users to stake \textit{Ether} and become validators since validators can earn huge profits from the \pri \txs. 
However, the difference of income between validators who are responsible to attest the blocks and the validators who are randomly selected to create blocks will increase significantly. 
In particular, users that have a large validator stake are likely to have more blocks to create and obtain more rewards from the MEV related \pri \txs. Thus, they earn more and have more to stake. 
Therefore, the impact of \pri \txs on Ethereum 2.0 is worth investigating, and Ethereum 2.0 will also affect the \pri \txs. 

\bheading{Private \txs on other blockchains.}
Similar as Ethereum, Binance Smart Chain (BSC) is built based on Ethereum Virtual Machine (EVM) and smart contracts. However, BSC uses a consensus of Proof of Staked Authority (PoSA)~\cite{bsc}. In BSC, there are only 21 validators that are similar as miners in Ethereum. Although the validators are re-elected daily based on the top 21 nodes that meet the requirements to be validators, the number is always 21. 
To the best of our knowledge, there is no official \pri \tx concept in the BSC while there are MEV Bots. We believe the reason is that it is hard to become validators in BSC than Ethereum miners since the there only be 21 validators. Besides, there are no special channels in BSC (e.g., FlashBots~\cite{flashbot} in Ethereum) for \pri \txs. 
For other blockchains, Polygon that can be considered as a fork of Ethereum uses a Proof-of-Stake (PoS) consensus mechanism. In order to participate in the blockchain, validators need to stake money and take the resposiblities as miners in Ethereum. Similarly, there are no general channels used for \pri \txs, although the MEV related \txs are popular in Polygan. 
We believe that there are more to observe and examine in such blockchains about MEV and \pri \txs. 

\section{Related Work}
\label{sec:related}
In this section, we discuss existing works that are related to this paper.
In particular, we first discuss existing works that cover private \txs,
then present the measurement papers on blockchains.
We also briefly discuss works that focus on finding bugs in smart contracts.

\subsection{Private Transactions}
\begin{figure}[t]
\centering
\includegraphics[width=0.65\columnwidth]{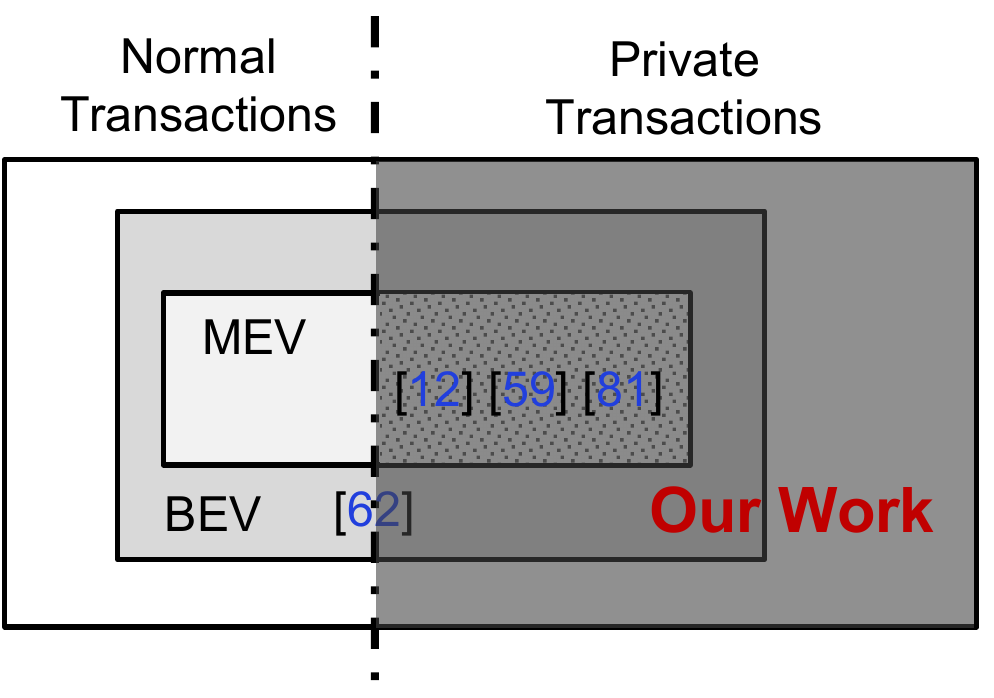}
\caption{The comparison between related works on \pri \txs and our work.
Piet~\etal~\cite{piet2022extracting}, 
Weintraub~\etal~\cite{weintraub2022flash}
and Capponi~\etal~\cite{capponi2022evolution}
focus on MEV in private \txs;
Qin~\etal~\cite{qin2021quantifying} investigate BEV, 
which is a super set of MEV;
Our work focuses on the entire private \tx landscape. }
\label{figs:related-works}
\vspace{-12pt}
\end{figure}

Private \txs are still very new and have not received much attention in the research community.
To date, there are only a few works~\cite{piet2022extracting, weintraub2022flash, capponi2022evolution, qin2021quantifying} that touch upon \pri \txs as by-products when investigating MEVs. We compare the scope of our work and these papers in~\figref{figs:related-works}. In particular, our work mainly study the landscape of \pri \txs, while they focus on MEVs.

\textit{Piet et al.}~\cite{piet2022extracting} measured \pri \txs in transaction pools from four sources for about 1 day, and found that most of the \pri \txs are used for miner profit redistribution and MEV extraction. In particular, 91.5\%  of MEV extraction is done via \pri \txs. They also analyzed the miner profits among \pri \txs and the top five miners earned most profits. 
Besides, 40\% of miners had not mined any \pri \tx. 
We are different in the focuses of analysis.
For example, they measure the inconsistency of \pri \txs in miners, while our work focuses on the profits flows earned by miners. 
Besides, we have a much larger dataset (one year) and we present a study on the security impacts of private \txs.  \looseness=-1

\textit{Weintraub et al.}~\cite{weintraub2022flash} mainly measure and analyze the impact of Flashbots with focusing on MEV, which provides a private channel between Ethereum users and miners. \textit{Weintraub et al.} also measures other private services providers (e.g., Eden Network~\cite{edennetwork}) and shows that most of the MEV extraction comes from private service providers, especially Flashbots. They discover the unfairness of Flashbots, in which miners have profited much more than MEV searchers.
Their work studies \pri \txs mainly from the perspective of MEV extraction, while our work gives a more complete view on \pri \txs in other aspects.

\textit{Qin et al.}~\cite{qin2021quantifying} measure the Blockchain Extraction Value (BEV) by detecting sandwich attacks, liquidation, and arbitrage in \txs. Specifically, they consider the \tx with zero gas price as \pri \txs and measure the percentage and its value in each BEV category. Moreover, their paper formalizes BEV relay systems and analyzes the impacts on P2P network and consensus layer. their paper mainly studies BEV, while our work highlights \pri \txs. 

\textit{Capponi et al.}~\cite{capponi2022evolution} propose  a game theoretical model to analyze the economic incentives behind the venues from the \pri \txs. In particular, their paper collected and investigated the transactions summited via the private channel from the Flashbots. The results show that the implications on different entities, including the increasement of users and miners payoffs. Their work highlights the economic incentives of \pri \txs, while our work focuses on analyzing the behavior of \pri \txs and their impacts.

\subsection{Measurement of Blockchains}
\bheading{Measurement of Ethereum interaction networks.}
To study the behaviors of different entities and their interactions, many works study Ethereum networks between users and smart contracts. Particularly, \textit{Lee et al.}~\cite{lee-measure}, ~\textit{Chen et al.}~\cite{chen2020understanding}, ~\textit{Lin et al.}~\cite{Zhao2021TemporalAO}, and ~\textit{Bai et al.}~\cite{bai-temporal} perform a measurement study on these interaction networks to observe the activities of Ethereum and propose some insights,
All of these works measure the general features of the Ethereum interaction networks between users and contracts from graph analysis, while our paper focuses on the purposes and usages of \pri \txs from the interaction networks.

\bheading{Measurement of Ethereum \txs.} 
Besides, some works~\cite{said-tx, lin-tx, zanelatto2020transaction} study \txs via graph analysis to investigate the \txs behaviors. 
Specifically, \textit{Said et al.}~\cite{said-tx} analyze transaction behaviors, community structure and link prediction. ~\textit{Lin et al.}~\cite{lin-tx} model the \txs records as temporal weighted multi-digraphs. ~\textit{Zanelatto et al.}~\cite{zanelatto2020transaction} conduct a large-scale study based on 3-year Ethereum \txs with cryptocurrency. 
These works measure the overall features of \txs, while we focus on investigating the influences of \pri \txs and the related security issues.
\looseness=-1

\bheading{Measurement and analysis of Ethereum EIP-1559.}
Moreover, works~\cite{liu2022empirical, Leonardos-eip1559, D-eip1559} conduct an empirical study on the effects of the EIP-1559 protocol. 
\textit{Liu et al.}~\cite{liu2022empirical} lead an empirical study the effects of EIP-1559 protocol on transaction fee dynamics, transaction waiting time, and security exploits.
\textit{Stefanos et al.}~\cite{Leonardos-eip1559} study both theoretical and experimental analysis of the dynamics and stability properties of the EIP-1559 protocol.
\textit{Daniel et al.}~\cite{D-eip1559} study the fee market upgrade by leveraging the data from first month after the EIP-1559 protocol. They also propose an alternative rule to replace EIP-1559 and prove the effectiveness from the experimental results.

\bheading{Measurement and analysis of MEV.}
In addition, MEV is gaining attentions recently. 
\textit{Daian et al.}~\cite{flashboy} is the first to propose the concept of MEV and empirically shows the risks posed by MEV.
\textit{Qin et al.}~\cite{qin2021quantifying} 
is the first to introduce the concept of Blockchain Extractable Value (BEV) and measure the dangers of BEV. Specifically, this paper quantifies the value threshold at which BEV would encourage miners to fork the blockchain.
\textit{Wang et al.}~\cite{wang2021cyclic} conduct a measurement study on \textit{cyclic arbitrages} which is one type of the arbitrages on the \textit{Uniswap} platform. 
Moreover, arbitrage is also measure in some works. 
\textit{Zhou et al.}~\cite{zhou2021just} propose a framework to automatically construct profitable arbitrage transactions. 
They analyze the MEV or in terms of its risks to the Ethereum consensus layer, while our article measures the impact of the MEV from the related \pri \txs.  Also, these papers detect or simulate the arbitrage in normal \txs while our work detects arbitrage in \pri \txs.

\subsection{Bug Detection on Smart Contracts}
Due to the immunity of smart contracts and their various applications, attacks happen consistently and frequently on Ethreum blockchain, which can cause huge financial losses. 
To mitigate such attacks, many works~\cite{grossman2017online, luu2016making, nikolic2018finding, tsankov2018securify, krupp2018teether, mythril, manticore,feist2019slither,kalra2018zeus,brent2018vandal,torres2018osiris,frank2020ethbmc,tikhomirov2018smartcheck} study the vulnerabilities of smart contracts and on Ethereum.
They usually use static analysis approaches, such as symbolic execution, to uncover vulnerabilities and bugs in smart contracts.
Some other works~\cite{jiang2018contractfuzzer,grieco2020echidna,wustholz2020harvey,nguyen2020sfuzz,choi2021smartian,groce2021echidna,he2019learning,liu2018reguard,torres2021confuzzius,fu2019evmfuzzer} make use of fuzzing techniques to discover bugs in smart contracts.
In contrast, our work analyzes \pri \txs to understand their mechanisms, incentives, and security risks. 

\section{Conclusion}
\label{sec:conclude}
In this paper, we conduct the \textit{first} empirical study on Ethereum \pri \txs. 
First, we study the general characteristics of \pri \txs, including 
the distribution, categories, 
DeFi tokens and platforms, 
and the involved entities. 
Contrary to the original intention of protecting users,
many \pri \txs are used 
for MEV opportunities by MEV searchers.
Second,
we study the transaction cost and miner profits of \pri \txs.
Private \txs use less gas and bring huge profits to miners.
Finally, we evaluate security-related issues, including MEV, real-world attacks, consensus security, and the leakage of \pri \txs. 
The huge profits of \pri \txs might lead to undercutting attacks,
and \pri \txs have about 4.3\% chance of being leaked to the public mempool. 
This work sheds light on the \pri \tx ecosystem, 
and calls for more actions to protect users from \pri \txs.



\newpage
\bibliographystyle{IEEEtranS}
\bibliography{paper}

\appendix
\section{Arbitrage Detection}
\label{sec:arbitrage}

Arbitrage usually triggers several token exchanges in \txs, in which arbitrageurs intend to gain profits by making use of the differentials of the token price in various platforms. 
We first describe the steps to extract swaps, which are the basic elements of detecting arbitrages. Then we introduce two types of arbitrages and their detection algorithms. Lastly, we perform detection on our one-year dataset and analyze the detected results.

\subsection{Extracting Swaps}
\begin{figure*}
        \centering
        \begin{subfigure}[b]{0.25\textwidth}
            \centering
            \includegraphics[width=\textwidth]{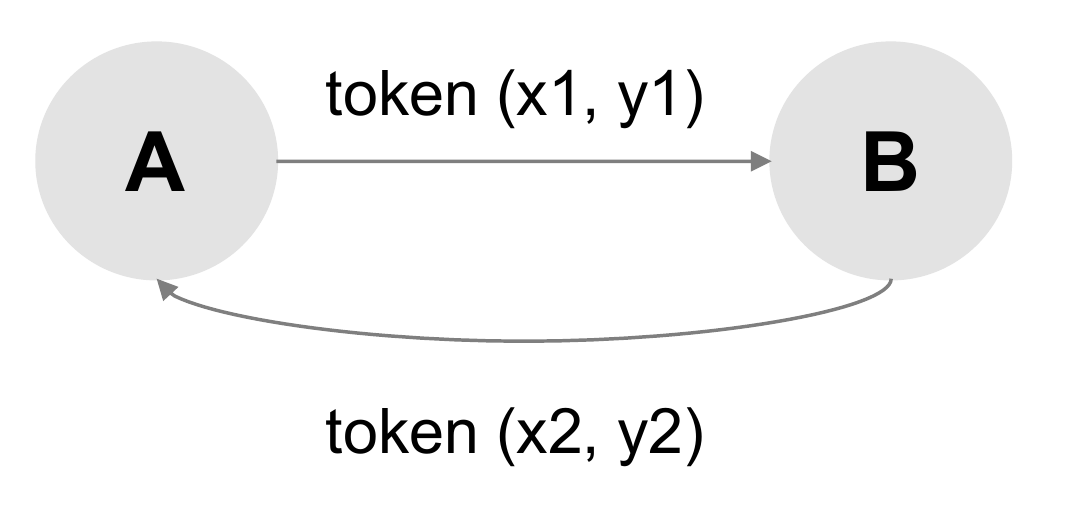}
            \caption[Network2]%
            {{\small The swap type I.}}    
            \label{figs:swap-type1}
        \end{subfigure}
        \hfill
        \begin{subfigure}[b]{0.675\textwidth}  
            \centering 
            \includegraphics[width=\textwidth]{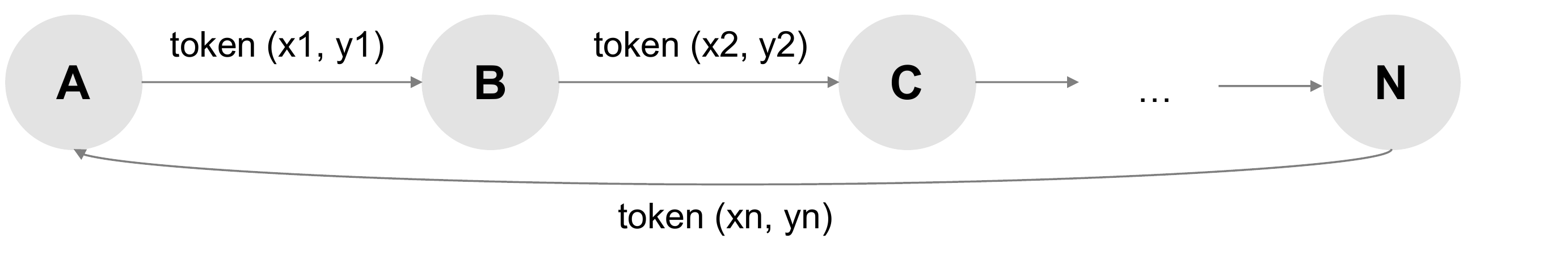}
            \caption[]%
            {{\small The swap type II.}}    
            \label{figs:swap-type2}
        \end{subfigure}
        \vskip\baselineskip
        \begin{subfigure}[b]{0.45\textwidth}   
            \centering 
            \includegraphics[width=\textwidth]{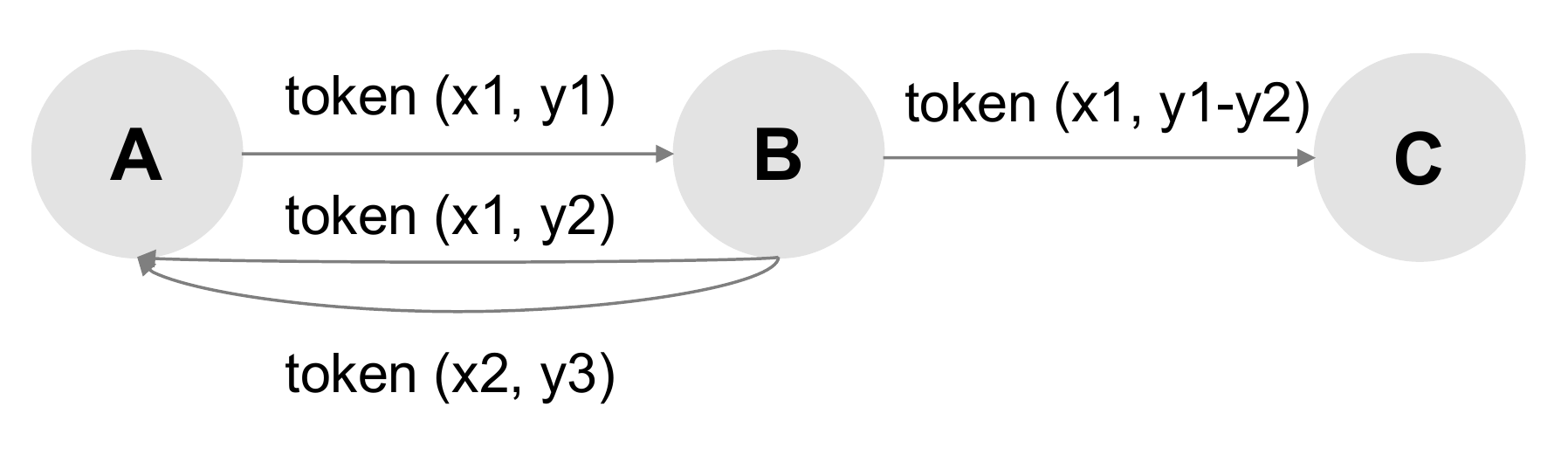}
            \caption[]%
            {{\small The swap type III.}}     
            \label{figs:swap-type31}
        \end{subfigure}
        \hfill
        \begin{subfigure}[b]{0.45\textwidth}   
            \centering 
            \includegraphics[width=\textwidth]{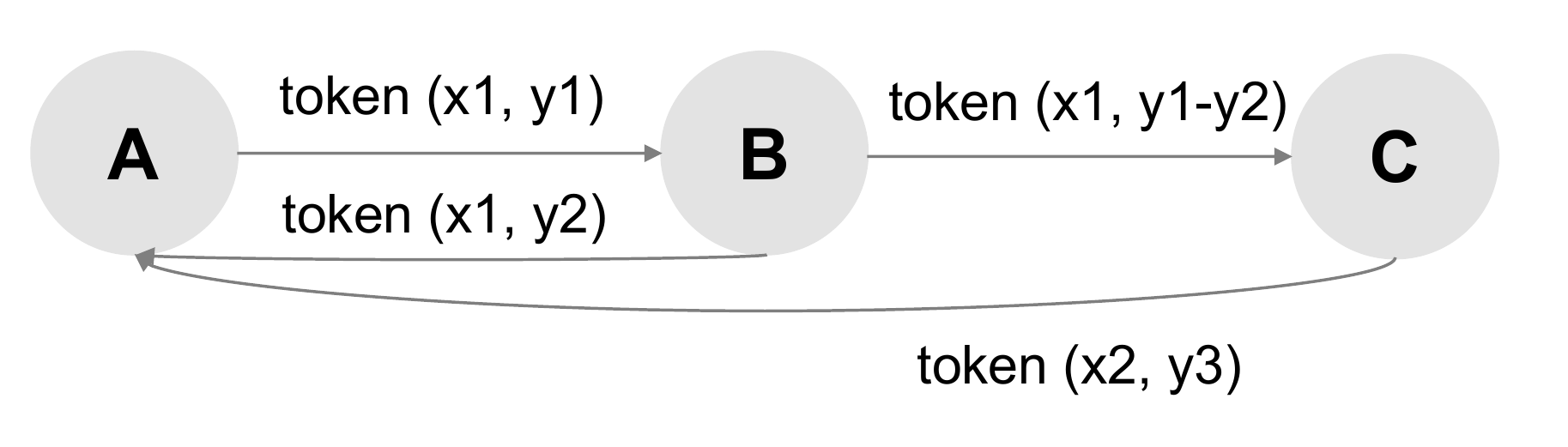}
            \caption[]%
            {{\small The swap type III.}}     
            \label{figs:swap-type32}
        \end{subfigure}
        \caption[ The average and standard deviation of critical parameters ]
        {\small Three types of swaps: x represents the token type; y means the amount of token; A, B, C, and N represent the addresses.} 
        \label{figs:swaps}
\end{figure*}

Swaps are the basic elements to detect arbitrage, which can be identified from the generated MFGs from \txs. In particular, we use the Johnathan's algorithm~\cite{Johnson1975FindingAT} to extract all the simple cycles in the MFG. Secondly, we exclude the unlikely cycles forming a swap and identify swaps based on their patterns. We classify the swaps into three types as follows and display the pattern of every type in~\figref{figs:swaps}. 
\begin{packeditemize}
\item {Type I:} \swap contains two token-exchange edges. These two edges form a cycle between nodes A and B. 

\item {Type II:} 
Similar to Type I, there should be a cycle; but the \swap involves more than two nodes, compared to Type I. There will be at least two different token types and three edges in the whole \swap. Usually, this type of \swap is started by node A which tries to exchange tokens with node B. But there are some intermediate nodes involved, and the process forms a cycle.

\item {Type III:} 
In such swaps, node A plans to exchange tokens with node B but B does not have enough tokens to transfer. Thus, the left tokens will be transferred back to A. Specifically, either node B or another node C which belongs to the same owners as B will send the tokens back to node A. 
Generally, there will be four edges in this type of swaps including two edges indicating the swap between node A and node B, one edge of sending the left token from B or C to node A, and another edge transferring the exchanged token from node B to node C.
\end{packeditemize}

\algtext*{EndIf}
\algtext*{EndFor}
\algtext*{Indent}

\begin{algorithm}[H] 
\label{alg:loop}
\begin{algorithmic}[1]
\caption{The algorithm of Cyclic Arbitrage}\label{alg:highslippage}
\Require{$Swaps$} 
\Ensure{$Cyclic$ $Arbitrage$ $Report$}
\Statex
\Function{CyclicArbitrage}{$Swaps$}
    \For{$Swap \gets Swaps_{1}$ to $Swaps_{N}$}   
        \If{$Swap$ is ordered}
            \State {$Ts$ $\gets$ {$Swap.Path$}}
            \For{$T \gets T_{1}$ to $T_{N}$}
                \If{$Check(T)$}
                    \State \Return {$CyclicArbitrageReport$}
                \EndIf
            \EndFor
        \EndIf
    \EndFor
\EndFunction

\end{algorithmic}
\end{algorithm}

\subsection{Cyclic Arbitrage}
Cyclic arbitrage has the token exchanges in a single swap consisting a cycle and should gain profits. 

\bheading{Detection Algorithm.}
To detect the cyclic arbitrage, we define our detection algorithm in ~\nalgref{alg:highslippage}. The algorithm first iterates the swaps one by one. For each iteration, the swap will be checked whether it is ordered. If the \textit{MFGIndex} in the swap is increasing one by one (e.g., [1, 2, 3]), the swap is ordered. With the ordered transfers in money flow, the balance of every Ethereum address will be calculated. If the address receives more tokens than sent, then it means the address launches the cyclic arbitrage and earns benefits via this transaction. These conditions are checked in the \textbf{Check} function.

\subsection{None-Cyclic Arbitrage}
None-Cyclic arbitrage has the token exchanges in several swaps, which can be detected by the following two requirements:
\begin{packeditemize}
    \item \textbf{Some swaps by the arbitrageur:} There should be some swaps in the transaction with different prices.
    \item \textbf{Arbitrageur makes money:} With all the swaps, the arbitrageur makes money that it receives more tokens than sent.
\end{packeditemize}

\bheading{Detection Algorithm.}
We define our detection rules based on the requirements in ~\nalgref{alg:arbitrage}. The algorithm first sorts the obtained swaps by the swap address. If an address contains two or more token swaps with different token pairs, then select it as the arbitrageur candidate since the none-cyclic arbitrage requires multiple swaps for the same token. Then the swaps will be looped and the balance of each involved Ethereum address will be calculated. If at least one address has the balance larger than 0, then it should be the arbitrageur and gains money.

\begin{algorithm}[t] 
\label{alg:loop}
\begin{algorithmic}[1]
\caption{The algorithm of None-Cyclic Arbitrage}\label{alg:arbitrage}
\Require{$Swaps$} 
\Ensure{$None-Cyclic$ $Arbitrage$ $Report$}
\Statex
\Function{NoneCyclicArbitrage}{$Swaps$}
    \State {$AddrSwaps_{N}$ $\gets$ {$SwapByAddr(Swaps)$}}
    \For{$AddrSwaps \gets AddrSwaps_{1}$ to $AddrSwaps_{N}$} 
        \If{$Check(AddrSwaps)$} 
            \If{ $Balance(AddrSwaps)>0$ } 
                \State \Return {$NoneCyclicArbitrageReport$}
            \EndIf
        \EndIf
    \EndFor
\EndFunction

\end{algorithmic}
\end{algorithm}


\subsection{Analysis}
Arbitrage is one type of MEV, which is pretty popular in rencent days. Arbitragers can submit \txs searching for opportunities and might cause financial losses to users. 
We detect these two types of arbitrages in 7,405,835 \pri \txs from the one-year dataset. Specifically, there are 1,099,006 \pri \txs flagged as arbitrage with a total value of approximately 389 million dollars. There are 376,393 cyclic and 722,613 non-cyclic arbitrages, respectively. 

     \begin{table}[h]
    \begin{center}
    \setlength{\tabcolsep}{4pt}
    \begin{tabular}{ lrrr} 
    \hline
    \scriptsize \textbf{Arbitrageur address} & \scriptsize \textbf{Flagged} & \scriptsize \textbf{Cyclic} &  \scriptsize \textbf{ Non-cyclic} \\ 
    \hline
    \scriptsize  0x74de5d4fcbf63e00296fd95d33236b9794016631 & \scriptsize 62,122 &  \scriptsize 62,121 &  \scriptsize 1 \\
     \scriptsize 0xdfee68a9adb981cd08699891a11cabe10f25ec44 &  \scriptsize 52,982 & \scriptsize  \scriptsize  0 & \scriptsize 52,982 \\ 
     \scriptsize 0x911605012f87a3017322c81fcb4c90ada7c09116 &  \scriptsize 37,968 &  \scriptsize 7,814 &  \scriptsize 30,154 \\
     \scriptsize 0x5aa3393e361c2eb342408559309b3e873cd876d6 &  \scriptsize 33,826 &  \scriptsize 0 &  \scriptsize 33,826 \\
     \scriptsize 0xa1006d0051a35b0000f961a8000000009ea8d2db &  \scriptsize 28,390 &  \scriptsize 14,184 &  \scriptsize 14,206 \\
     \scriptsize 0x5144aca278867ab6c04455464d8f6a2607e3c075 &  \scriptsize 26,339 &  \scriptsize 18,028 & \scriptsize  8,311 \\
     \scriptsize 0x9008d19f58aabd9ed0d60971565aa8510560ab41 &  \scriptsize 24,297 &  \scriptsize 24,285 &  \scriptsize 12 \\
    \scriptsize  0x000000000000abe945c436595ce765a8a261317b &  \scriptsize 24,065 &  \scriptsize 6,325 &  \scriptsize 17,740 \\
     \scriptsize 0x018d5c4783f5317815f6e8168942a12adde3cd3c &  \scriptsize 23,869 &  \scriptsize 4,926 &  \scriptsize 18,943 \\
     \scriptsize 0x7cf09d7a9a74f746edcb06949b9d64bcd9d1604f &  \scriptsize 22,158 &  \scriptsize 1,506 &  \scriptsize 20,652 \\
    \hline
    \end{tabular}
    \caption{The top 10 arbitrageurs in \pri \txs.}
    \label{tab:top_Arbitraguers}
    \end{center}
   \end{table} 
   
    \begin{figure}[t]
    \centering
    \includegraphics[width=1\columnwidth]{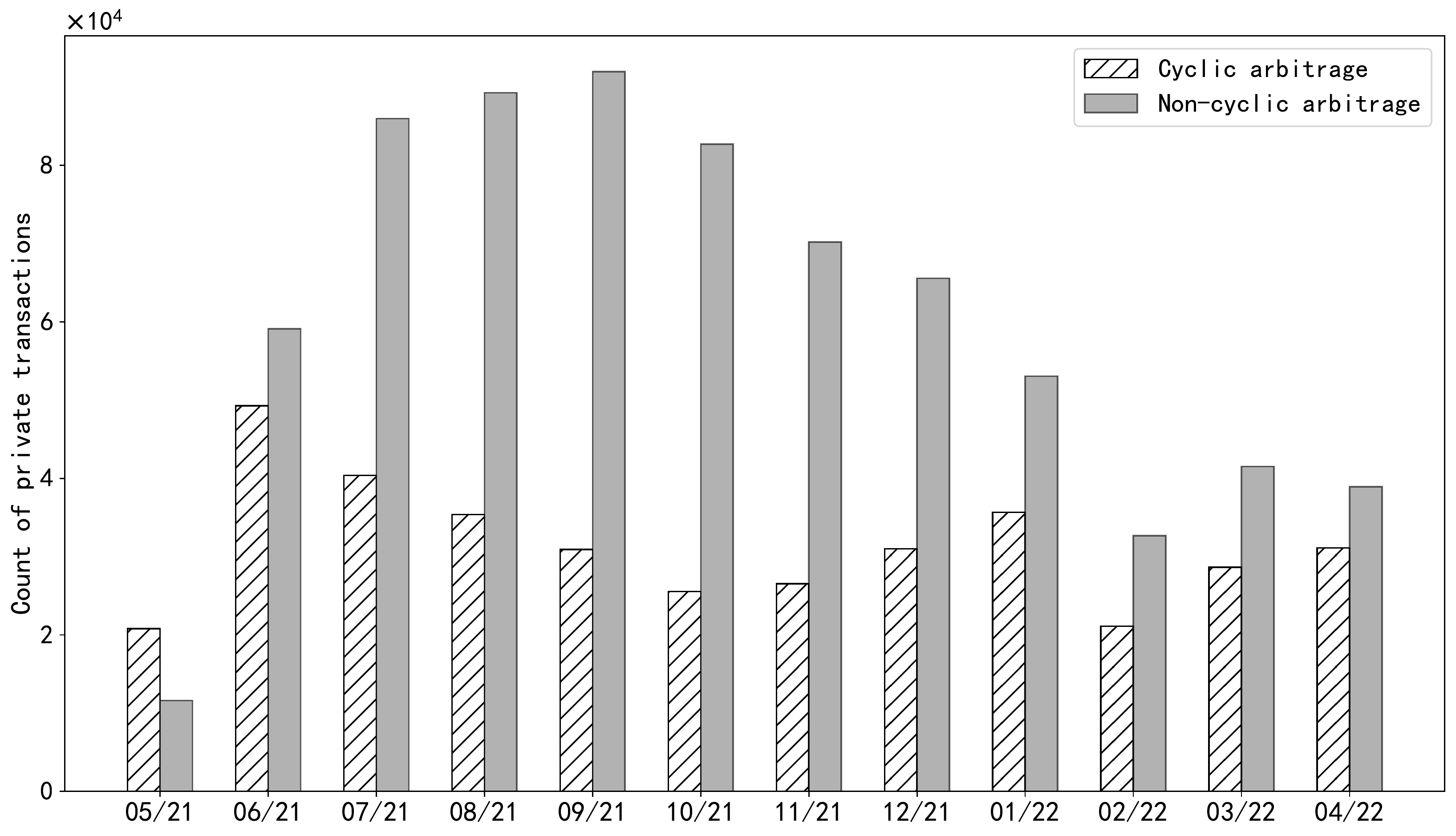}
    \caption{The count of arbitrage in \pri \txs per type by month.}
    \label{figs:arb-type}
    \end{figure}

    \bheading{Top 10 arbitraguers in \pri \txs.}
    We identify 1,456 arbitrageurs and present the top 10 arbitrageurs in~\tabref{tab:top_Arbitraguers}. The arbitrage \txs in the top 10 arbitrageurs account for 31\% of the total, and the cumulative profits take around 10\%. Most of the top arbitrageurs have preferences on the type of arbitrages. For example, the top 1 arbitrageur only has 1 non-cyclic arbitrage \tx among the flagged 62,122 arbitrage \txs. Moreover, 95\% of the arbitrage \txs occur in popular DeFis, including Uniswap~\cite{uniswap-v3}, Sushiswap~\cite{sushiswap}, and linch~\cite{1inch}. 
    Besides, we present the count of two types of arbitrage per month in~\figref{figs:arb-type}. In almost every month, there are more none-cyclic than cyclic arbitrage \pri \txs. In particular, from July 2021 to December 2021, the count of none-cyclic arbitrage \pri \txs is two or three times of the cyclic arbitrage.

    \begin{figure}[t]
    \centering
    \includegraphics[width=1\columnwidth]{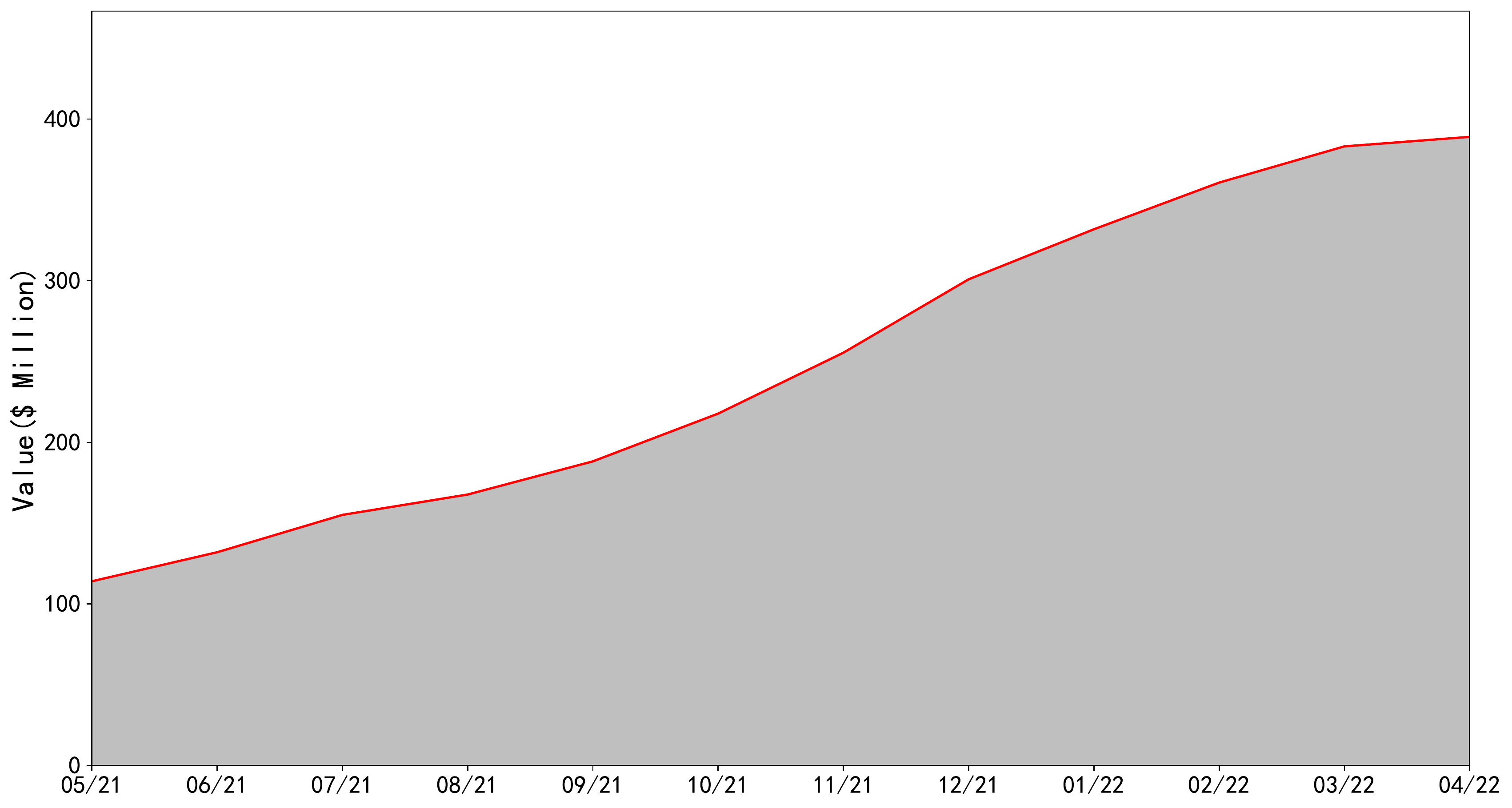}
    \caption{The cumulative profit of arbitrage in \pri \txs by month.}
    \label{figs:trec}
    \end{figure}
    
    \bheading{Profit of the arbitrage \pri \txs.}
    We present the profits of arbitrage by month in~\figref{figs:trec}.
    The total profits earned from all arbitrage \txs is about 389 million dollars, of which the profit (around 1.1 million dollars) in May 2021 is the highest. Moreover, about 40\% of the profits flow to the miner through transaction fees and direct payments to miners.
    
    \bheading{Position of the arbitrage \pri \txs in blocks.}
    About 97\% of the arbitrage \pri \txs are within the top 10 inside the blocks. For the rest 3\% arbitrages that are outside the top 10, their mined blocks usually have lots of \txs and \txs in the front positions inside these blocks usually transfer profits to miners, which causes huge competitions. It is the reason why arbitrage \pri \txs cannot be placed in the very front of blocks.
    
    \bheading{Tokens in arbitrage \pri \txs.} 
    We present the count of arbitrage \pri \txs with the specific number of token types and DeFis in~\tabref{tab:tokens}, which are used for exchanging tokens. 
    Arbitrageurs usually make use of the differentials of token prices in different DeFis, to earn profits. From the table, arbitrage \pri \txs involving in 2 DeFis accounts for the most. Moreover, we observe that arbitrageurs normally will hold stablecoins (e.g., USDT, USDC) as the profits in the end of the arbitrage \pri \txs. 
    
    \begin{table}[h]
    \begin{center}
    \begin{tabular}{c|ccccc} 
    \hline
     \scriptsize \diagbox{Token Type}{DeFi} & \scriptsize  1 &  \scriptsize 2 &  \scriptsize 3 & \scriptsize  4 &  \scriptsize $\geq{5}$ \\
    \hline
    \scriptsize 2 & \scriptsize 19,586 &  \scriptsize 326,445 &  \scriptsize 116,071 & \scriptsize  8,645 &  \scriptsize 387 \\
     \scriptsize 3 &  \scriptsize 205,243 &  \scriptsize 217,684 &  \scriptsize 57,802 &  \scriptsize 66,142 &  \scriptsize 6,416 \\
    \scriptsize  4 &  \scriptsize 84,451 & \scriptsize  22,187 &  \scriptsize 28,513 & \scriptsize  9,033 &  \scriptsize 8,198 \\
     \scriptsize $\geq{5}$ &  \scriptsize 15,850 &  \scriptsize 3,130 &  \scriptsize 6,598 &  \scriptsize 1,946 &  \scriptsize 10,750 \\
    \hline
    \end{tabular}
    \caption{Token swap categories of arbitrage \txs}
    \label{tab:tokens}
    \end{center}
   \end{table} 

\end{document}